\documentclass[onecolumn]{aa} 

\usepackage{soul}
\usepackage{graphicx}
\usepackage{xcolor}
\usepackage{booktabs} 
\usepackage{multirow} 
\usepackage{ulem}

\newcommand{\ii}{{\sc ii}}
\usepackage{txfonts}
\usepackage[draft]{hyperref}

\begin{document}

   \title{
   Long-term activity cycles in planetary M stars observed with SOPHIE}
\titlerunning{Activity cycles in M stars}

   \author{C. G. Oviedo
          \inst{1,2},
          A. P. Buccino\inst{1,2},
          R.F. Díaz\inst{3,4}
          R. Petrucci\inst{2,5},
          E. Jofré\inst{2,5},
          I. Boisse\inst{6},
          P. D. Colombo\inst{1,2}
          \& X. Delfosse\inst{7}
          }

\institute{Instituto de Astronomía y Física del Espacio, CONICET--UBA, Argentina,
\and
              Consejo Nacional de Investigaciones Científicas y Técnicas (CONICET), Godoy Cruz 2290, CABA, CPC 1425FQB, Argentina,
\and
              Instituto Tecnológico de Buenos Aires (ITBA), Iguazú 341, Buenos Aires, CABA C1437, Argentina,
              \and
Instituto de Ciencias Físicas (CONICET / ECyT-UNSAM), Campus Miguelete, 25 de Mayo y Francia, (1650) Buenos Aires, Argentina,
\and
              Universidad Nacional de Córdoba - Observatorio Astronómico de Córdoba, Laprida 854, X5000BGR, Córdoba, Argentina,
              \and
              Aix Marseille Univ, CNRS, CNES, LAM, Marseille,
France
              \and
              Univ. Grenoble Alpes, CNRS, IPAG, 38000 Grenoble, France\\
              \email{coviedo@iafe.uba.ar}
}

\abstract 
{M dwarfs are prime targets for exoplanet searches due to their low masses and radii, which enable the detection of small planets in their habitable zones (HZs). However, the magnetic activity of M dwarfs can introduce signals in radial velocity measurements that may be mistaken for planetary signatures, making the understanding of stellar activity cycles crucial for accurate planet detection and characterisation.} 
{We aim to identify and characterise long-term magnetic activity cycles in M dwarfs using a homogeneous and extensive spectroscopic dataset in order to better understand their magnetic variability and its implications for exoplanet detection.} 
{We analysed 13 years of high-resolution spectra obtained with the SOPHIE spectrograph for two early M dwarfs known to host exoplanets. We simultaneously monitored chromospheric activity using two indicators, the H$\alpha$ index and the Mount Wilson $S$-index. Long-term trends were modelled using both sinusoidal and low-order polynomial fits to robustly identify stellar activity cycles. As a complement, we used TESS photometric data to assess the short-term variability of both targets.} 
{We detected long-term variability consistent with stellar magnetic cycles in both targets. For GJ 617A, we report a cycle of approximately 4.8 years, while for GJ 411, we find several characteristic timescales of variability of about 4.9 years. In addition, TESS photometric data reveal signs of short-term variability in GJ617A.} 
{The periods of the long-term variability detected for GJ 617A and GJ 411 do not coincide with any of the planetary signals previously reported, which reinforces the hypothesis that they are of magnetic origin. If indeed the variability is due to activity, the cycles detected would not be driven by the same mechanism: The cycle in GJ 617A is consistent with a solar-like dynamo, while the rotation seems to play a different role in the long-term cycles detected in GJ 411.} 


\maketitle
%

\section{Introduction}

M-type stars ($M_{\ast} \leq 0.6 \, M_{\odot}$) are the most abundant in the solar neighbourhood ($d$\ <\,10\,pc; \citealt{Bochanski2010}). They account for approximately 80\% of the total stellar population in the Galaxy (RECONS Survey; \citealt{RECONS2023}). 
Due to their low luminosity and the proximity of their habitable zone (HZ) compared to solar-type stars \citep{Turbet2023,Kopparapu2013}, M-type stars provide ideal conditions for detecting potentially habitable Earth-sized planets using radial-velocity (RV) and transit methods \citep{Snellen2015} despite the challenges posed by their frequent stellar activity. Indeed, a large fraction of M stars exhibit high chromospheric activity that exceeds that of the Sun and show frequent, highly energetic flares (e.g. \citealt{RodriguezMartinez2020,SuarezMascareno2016,SuarezMascareno2018,IbanezBustos2019b}).

The influence of magnetic activity and stellar rotation on RVs is linked to the presence of surface inhomogeneities, such as spots and plages, which traverse the stellar disc during rotation. These features distort the profiles of spectral lines, introducing spurious variations in RV measurements. Starspots, for instance, dim the light from the rotating stellar disc's edges, causing shifts in the centroid of spectral lines \citep{Saar1997}. Plages produce a similar effect, but with an opposite sign \citep{Meunier2010}. The rotation periods of field-age dwarf stars vary significantly across the lower main sequence, exceeding 100 days for mid-to-late M dwarfs \citep{Newton2016}. Within this same stellar mass range, the orbital periods of planets in the HZ decrease from approximately 365 days to as short as 10 days, resulting in an overlap between stellar rotation periods and the orbital periods of potentially habitable planets. Disentangling these signals from the reflex motion induced by exoplanets requires careful modelling of the variations in spectral line profiles \citep{Queloz2009,Boisse2011,Dumusque2011model,Rajpaul2015}. 

Clear and non-controversial examples of an exoplanet detection around an M dwarf have been firmly established in the literature, for example, Proxima b and GJ 667Cc \citep{AngladaEscude2016, Bonfils2013}. In parallel, stellar activity has been shown to mimic RV signals consistent with HZ planets, as illustrated by the case of Gliese 581, where the signals originally attributed to planets GJ 581d and GJ 581g were later reinterpreted as artefacts induced by stellar activity \citep{Robertson2014}.

Long-term variations in magnetic activity associated with stellar cycles are one of the primary limitations in identifying long-period planetary companions with the RV method. In solar-type stars, activity cycles last from a few years to more than a decade \citep{Baliunas1995, Lovis2011} and can induce RV signals with amplitudes of up to 10 m/s \citep{Lovis2011, Dumusque2011}. In addition, studying stellar cycles in M dwarfs is particularly challenging, as their low intrinsic luminosity requires longer exposure times or more sensitive instruments. In practice, this hinders the systematic acquisition of the long-term continuous time series needed to characterise their activity cycles. Despite these limitations, recent studies have identified long-term variations in M-type stars, for instance \citep{GomesdaSilva2011, GomesdaSilva2012, SuarezMascareno2018, Mignon2023, IbanezBustos2025, Buccino2011}. These efforts have revealed cycles with typical periodicities of 6.0 years in early M-class stars and 7.1 years in mid-M-class stars. \citep{SuarezMascareno2016}.

Therefore, if stellar cycle effects are not adequately corrected in RV data, spurious low-frequency signals can emerge, thus complicating the detection of planets with similar orbital periods. Furthermore, by interacting with the typically complex window function of exoplanet RV observations, the unaccounted for stellar activity cycle can contaminate periodogram analyses over a wide range of frequencies.

The Ca \ii\- H and K resonance lines (3969 \AA\-  and  3934 \AA) are key indicators for analyzing stellar activity, as they reflect variations in the chromosphere. These lines exhibit broad Ca+ absorption from the cooler upper photosphere and lower chromosphere along with narrow Ca+ emission from the hotter upper chromosphere. Since the 1960s, particularly through the Mount Wilson Observatory program \citep{Wilson1968}, the $S$-index has been used to quantify stellar chromospheric activity.
Although Ca \ii\- H and K lines have traditionally been used as chromospheric indicators in solar-type stars, they are not ideal for red dwarfs, where observing the Ca \ii\- lines requires long exposure times to obtain reliable observations. Significant efforts have been made to identify more suitable activity indicators in the redder wavelengths (e.g. Na {\sc{i}} in \cite{Diaz2007}, Ca \ii\- triplet in \cite{Martin17}, He {\sc{i}} in \cite{Fuhrmeister20}, and H$\alpha$ in \citealt{IbanezBustos23}).

For several decades, the H$\alpha$ line (6562.8 \AA) has played a crucial role in characterising the activity of M stars, as its presence in emission or absorption is used to distinguish between active and inactive M stars. This line traces the upper chromosphere, providing complementary information about chromospheric conditions (\citealt{Mauas1994}, \citealt{Mauas1996}; \citealt{Leenaarts2012}). Recent studies have highlighted the relevance of the H$\alpha$ index for understanding stellar activity in M dwarfs, showing that $\log_{10} \left({L_{H\alpha}}/{L_{\text{bol}}}\right)$, which quantifies the emission in the  H$\alpha$ line normalised to bolometric flux, is significantly correlated with the stellar rotation period \citep{Newton2017}. This reinforces the utility of H$\alpha$ as a stellar activity indicator, especially in studies of late-type stars. However, the use of the H$\alpha$ line as a long-term chromospheric activity indicator has been under debate in the literature  \citep{GomesDaSilva2014, Flores2018, GomesdaSilva22}, as the correlation between Ca \ii\- and H$\alpha$ seems to not be straight (e.g. \citealt{Cincunegui2007}). Recently, \cite{Meunier2024} concluded that the relation between  Na {\sc{i}}, Ca \ii\- and H$\alpha$ indices presents a large diversity in behaviour over a particular sample of 177 M dwarfs. 

In addition to spectroscopic and photometric approaches, high-resolution polarimetry has emerged as a key tool for detecting magnetic cycles in M dwarfs. Recent studies using SPIRou and ESPaDOnS have tracked long-term variations in the large-scale magnetic field topology of several stars, including polarity reversals and secular evolution (e.g.\ \citealt{Donati2023, Lehmann2024}).

In this work, we present a systematic study of long-term stellar activity in the M dwarf stars GJ 617A and GJ 411 using data from the Spectrographe pour l’Observation des Phénomènes des Intérieurs stellaires et des Exoplanètes (SOPHIE spectrograph).
To complement the long-term spectroscopic analysis, we also investigated high-cadence photometric data from the Transiting Exoplanet Survey Satellite (TESS) satellite covering multiple sectors for both stars and assessed their Galactic kinematic membership using astrometric data from the Gaia DR3 catalogue.

This paper is structured as follows. In Sect. 2 we present the stellar sample along with its main properties, and we describe the spectroscopic and photometric datasets used in this study. In Sect. 3 we describe the methodology employed to trace long-term stellar activity using spectroscopic indicators. Section 4 presents our main results, including the detection of long-term activity signals and the correlation between the studied indicators.
We detail the search for short-term photometric variability based on TESS observations in Sect. 5. Section 6 is dedicated to the analysis of the Galactic kinematic membership of the two targets. We also discuss the Galactic context of both stars based on their kinematics. Finally, our conclusions are summarised in Sect. 7.

\section{Observations and stellar sample}

\subsection{High-resolution spectroscopy with SOPHIE}

The SOPHIE spectrograph is a high-resolution (R=75000) cross-dispersed spectrograph mounted on the 1.93-metre telescope at the Haute-Provence Observatory in southern France \citep{Perruchot2008}. It covers the optical range of 3872–6943 Å and has been employed for a variety of astrophysical studies. In particular, SOPHIE enables the exploration of stellar activity cycles and the distinction of signals associated with potential planets in M dwarfs \citep{Perruchot2008, Bouchy2009}. SOPHIE's broad spectral range also allows for detailed characterisation of stellar activity indicators spanning from the lower to the upper chromosphere. This capability is especially valuable for long-term studies of stellar activity cycles in M dwarf stars.

The instrument design incorporates a fibre link from the Cassegrain focus of the telescope to ensure high-precision RV measurements, with an accuracy of 2–3 m/s for late-type stars (F, G, K, and M), depending on the signal-to-noise ratio (S/N) of the spectra \citep{Perruchot2008, Bouchy2011}. In 2011, SOPHIE was upgraded to SOPHIE+ through the introduction of an octagonal-section fibre in the fibre link, which significantly improved the stability of RV measurements \citep{Perruchot11, Bouchy13}. Thanks to this instrumental upgrade, combined with improvements in the data-reduction software, SOPHIE has since achieved a precision of 1–2 m/s for bright stars. The accuracy and consistency of its measurements are further ensured by recording the spectrum of a reference lamp together with the stellar spectrum, which allows for the correction of instrumental drift \citep{Bouchy2011}.

In recent years, SOPHIE has continued to deliver high-impact science. Examples include the discovery of three warm Jupiters in combination with TESS and CHEOPS \citep{heidari2024}, the detection of a new circumbinary planet (BEBOP-3b; \citealt{Baycroft2025}), and the joint identification of a super-Earth candidate around Gl 725A with SPIRou \citep{CortesZuleta2024}.

{\subsection{Data selection}}

\noindent The SOPHIE database consists of thousands of high-resolution optical spectra covering the spectral range of lines used to calculate various stellar activity indicators (such as the Ca {\sc ii}, Na {\sc i} D, and H$\alpha$ lines) and spanning a period of nearly two decades. Therefore, the SOPHIE database provides an ideal dataset for conducting long-term activity analyses of cool stars.

\begin{figure}[htb!] 
    \centering
    \includegraphics[width=1.0\columnwidth]{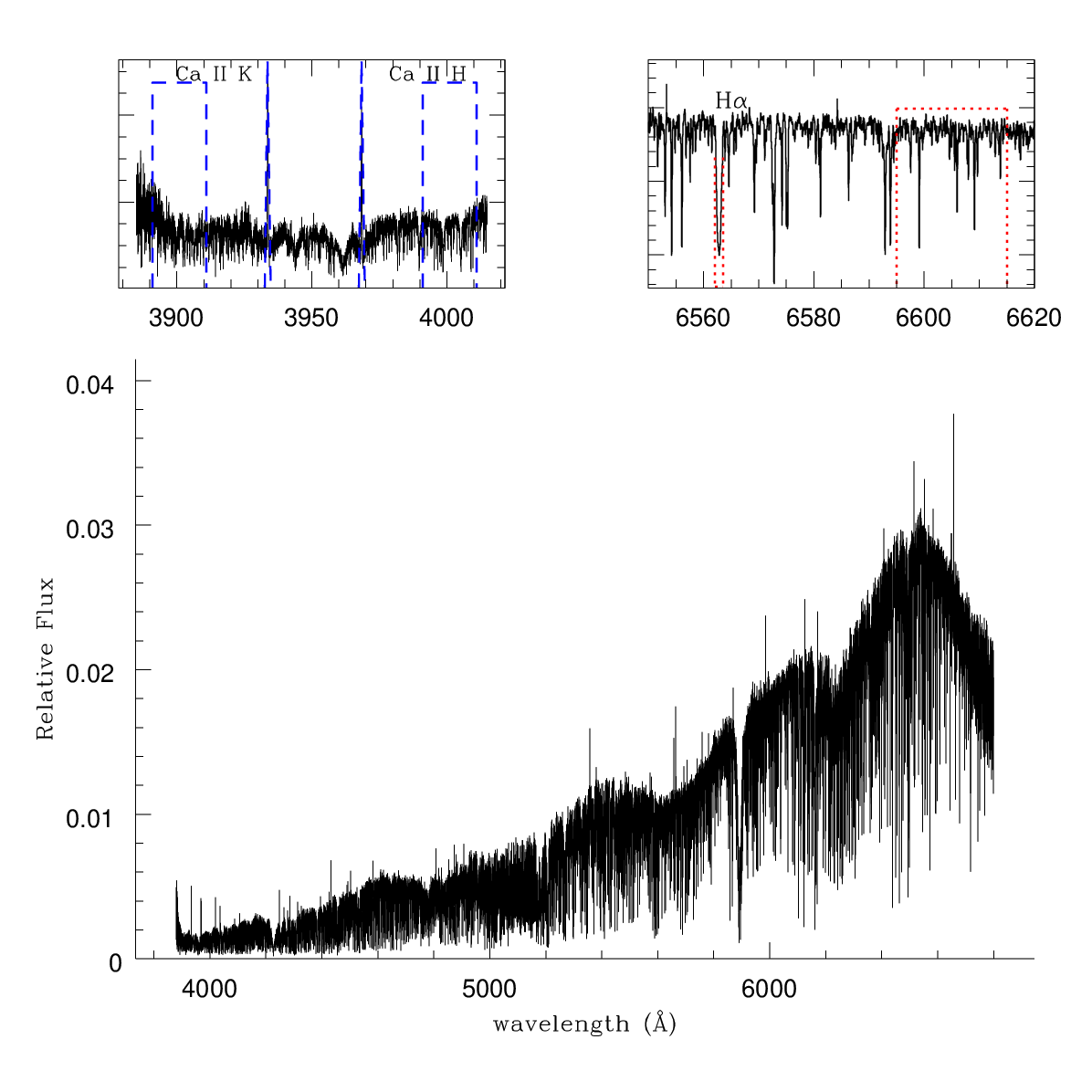}
    \caption{Spectrum from the SOPHIE database for GJ 617A. A zoomed-in view highlights the Ca {\sc ii} K (in emission) with a dashed blue line and the H$\alpha$ lines with a dashed red line, along with the H$\alpha$ region.}
    \label{Figura1}
\end{figure}

A total of 525 M-type stars were observed in high-resolution mode with the SOPHIE spectrograph, resulting in 7,695 publicly available spectra. A large fraction of these were obtained by the SOPHIE Consortium \citep{Bouchy2009}, which is dedicated to the search and characterisation of extrasolar planets around different types of stars (e.g. \citealt{hebrard2010, boisse2012, moutou2014, diaz2016, Hobson2018, demangeon2021, hara2020, heidari2024}).
These data were processed using the SOPHIE standard reduction pipeline \citep{Bouchy2009}, which produces s1d spectra by re-grouping and re-connecting the individual spectral orders. 
In particular, we selected the stars GJ 617A and GJ 411 for their extensive observational datasets.
The SOPHIE observations span from 2008 to 2024 for GJ 617A and from 2007 to 2024 for GJ 411, providing a sufficiently long baseline to investigate stellar magnetic activity on long timescales.
In Fig.~\ref{Figura1} we present a SOPHIE spectrum of an M-type star, highlighting the Ca {\sc ii} K and H lines, along with the H$\alpha$ line.

The first stage of our methodology focused on evaluating and filtering the spectra based on their signal-to-noise ratio (S/N). To ensure reliable measurements of the activity indices, specific S/N thresholds were defined at the wavelengths relevant to each indicator. In particular, we retained only spectra with S/N > 50 measured in order 38, which contains the Ca II H and K lines. As a result of this filtering, approximately 45 spectra were discarded for GJ 617A and about 15 spectra for GJ 411.

In addition to the spectral quality criterion, we focused on selecting stars with time series long enough to allow the detection of robust stellar cycles. Following the statistics presented by \cite{SuarezMascareno2016}, we adopted a minimum observational baseline of seven years. This criterion allowed us to align our sample selection with the known distribution of activity cycles in M dwarfs \cite{SuarezMascareno2018}, ensuring that the longer stellar periods typical of this type of star can be reliably detected.

The SOPHIE observations analysed in this work span 2008–2024 for GJ 617A and 2007–2024 for GJ 411, providing a sufficiently long temporal baseline to investigate stellar magnetic activity on long timescales.
\\

\section{Long-term activity analysis}  

To detect long-term activity cycles or systematic variations in M dwarf stars, both with and without planetary companions, we first constructed time series of activity proxies. Taking advantage of the broad wavelength coverage of SOPHIE spectra, we analysed simultaneous measurements of H$\alpha$ and Ca \ii\- stellar activity indices. To do so, we computed dimensionless activity indices in both lines following \cite{Cincunegui2007}. The Ca \ii\- index, hereafter $S$-index, is computed as the ratio between the mean H and K line-core fluxes integrated with a  triangular profile with a full width at half maximum of 1.09 Å (see Fig. \ref{Figura1}) to the mean continuum nearby integrated over two windows of 20 Å centred at m 4001 and 3901 Å. The H$\alpha$-index is defined as the ratio of the average line-core flux integrated with a rectangular band of a width of 0.7 Å (see Fig. \ref{Figura1}) centred at 6592.8 Å and the average flux or number of counts in the continuum nearby centred at 6605 Å and with a width of 20 Å.

To clean the time series of the activity indices from any outliers or flares (Fig. \ref{flares}; see Sect.~\ref{sec:flares} for more details) that might affect the accuracy of the results, we first computed a moving mean with a 90-day window. Outlier detection was based on a 2$\sigma$ threshold from this smoothed curve. In cases where outliers were identified, they were removed if their values were significantly above the typical range of the series, thus allowing the analysis to focus on stellar activity associated with long-term phenomena, such as starspots or faculae. To preserve the temporal coherence of the datasets, all measurements rejected from the time series of a given activity index were also removed from the time series of the other index.

A total of 480 spectra were available for GJ 617A, of which 23 were identified as outliers and excluded, resulting in 457 spectra used in the final analysis. For GJ 411, 341 spectra were available, with 57 identified and excluded as outliers, leaving 284 spectra for the final time series.

Finally, we proceeded with the analysis aimed at identifying long-term stellar activity cycles. To this end, we applied two complementary approaches based on the standard generalised Lomb--Scargle (GLS) periodogram \citep{Zechmeister09}. The classical GLS models the data using a purely sinusoidal basis and assumes a constant baseline, an assumption that may not hold in the presence of slow non-sinusoidal variability or long-term trends.

To relax this constraint, we adopted an extended formulation of the GLS, hereafter referred to as LinGLS (linear generalised Lomb--Scargle), which augments the harmonic basis with a low-order polynomial component. This approach allows simultaneous modelling of periodic signals and long-term variations, and it is particularly well suited for irregularly sampled time series affected by secular trends.

We denote by ${(t_i, y_i)}_{i=1}^{N}$ the observation times and corresponding chromospheric activity measurements, with associated uncertainties $\sigma_i$. A constant uncertainty was adopted for all data points, justified by the instrumental stability of the activity index. Long-term variations were modelled using a second-order polynomial basis,
\begin{equation}
\Phi_{\mathrm{poly}}(t_i)
=
\begin{bmatrix}
1 \\
t_i - \bar{t} \\
(t_i - \bar{t})^2
\end{bmatrix},
\end{equation}
where $\bar{t}$ is the mean observation time.

An initial fit using only the polynomial component yielded a reference model and a reduced chi-square, $\chi^2_0$. For each trial frequency, $\nu$, drawn from a densely sampled frequency grid defined by the temporal baseline and an oversampling factor, the design matrix was extended to include sinusoidal terms:
\begin{equation}
\Phi_{\nu}(t_i)
=
\left[
\Phi_{\mathrm{poly}}(t_i),
\;
\sin(2\pi \nu t_i),
\;
\cos(2\pi \nu t_i)
\right].
\end{equation}

The full model,
\begin{equation}
y(t_i) =
a_0 + a_1 (t_i - \bar{t}) + a_2 (t_i - \bar{t})^2
+ A \sin(2\pi \nu t_i) + B \cos(2\pi \nu t_i),
\end{equation}
was fitted by weighted linear least squares, and we solved the normal equations via Cholesky decomposition and included a small regularisation term of $10^{-8}$ to ensure numerical stability. The reduced chi-square $\chi^2(\nu)$ was computed for each frequency.

The LinGLS power was defined as
\begin{equation}
P(\nu) = \frac{\chi^2_0 - \chi^2(\nu)}{\chi^2_0}
\end{equation}
such that high values of $P(\nu)$ indicate a significant improvement of the fit when including a periodic component at frequency $\nu$.

By explicitly modelling long-term trends, LinGLS avoids biases associated with a constant baseline assumption and reduces sensitivity to slow instrumental or astrophysical variations. This makes it particularly well suited for the analysis of chromospheric activity indices in M dwarfs, whose variability often departs from simple sinusoidal behaviour.

To further exploit the information provided by the LinGLS framework, we analysed the statistical significance of the polynomial coefficients associated with the dominant frequency identified in the periodograms. For both GJ 411 and GJ 617A, we performed an ordinary least-squares fit of a sinusoidal model combined with polynomial terms to the activity indices, fixing the frequency to that of the main periodogram peak.

All data points were assigned equal weights, which is equivalent to an unweighted fit. The statistical significance of each coefficient was then evaluated using Student’s t-test, and the results were summarised using standard significance levels.

We estimated the false alarm probabilities (FAPs) using a bootstrap re-sampling method. The FAP for an observed periodogram power, $P$, was determined using a bootstrap re-sampling approach. We generated 10,000 synthetic time series by resampling the original data (or residuals) under the null hypothesis of no periodicity. The periodogram was computed for each synthetic series. The FAP was then calculated as the fraction of these synthetic periodograms that showed a maximum power equal to or greater than the observed power, $P$.

\section{Magnetic activity cycles in GJ 617A and GJ 411}

Based on the GLS and LinGLS periodograms described above, we identified stars exhibiting clear periodic signals with a false-alarm probability (FAP) below 1\%. In other cases, the periodograms showed no significant periodicity or variability.

\begin{table}[ht!]
\caption{Key stellar parameters of GJ~617A and GJ~411.}
\label{Table1}
\centering
\begin{tabular}{lcc}
\hline\hline
Parameter & GJ 617A & GJ 411 \\
\hline
Spectral type & M0.0\,V$^{\text{a}}$ & M1.5\,V$^{\text{a}}$ \\
$V$ (mag) & 8.90$^{\text{b}}$ & 7.49$^{\text{b}}$ \\
Distance (pc) & $10.767 \pm 0.002^{\text{c}}$ & $2.544 \pm 0.002^{\text{c}}$ \\
$T_{\rm eff}$ (K) & 3770$^{\text{c}}$ & 3511$^{\text{c}}$ \\
$[\text{Fe}/\text{H}]$ (dex) & $0.16 \pm 0.16^{\text{d}}$ & $-0.36 \pm 0.08^{\text{e}}$ \\
$M$ ($M_\odot$) & $0.58 \pm 0.08^{\text{d}}$ & $0.38 \pm 0.01^{\text{e}}$ \\
$R$ ($R_\odot$) & $0.63^{\text{c}}$ & $0.40^{\text{c}}$ \\
$L$ ($L_\odot$) & 0.96$^{\text{c}}$ & $2.25 \times 10^{-6}{}^{\text{c}}$ \\
$P_{\rm rot}$ (d) & 22.0$^{\text{f}}$ & 56.15$^{\text{g}}$ \\
$\log R'_{\rm HK}$ & $-4.75 \pm 0.14^{\text{h}}$ & $-5.47 \pm 0.10^{\text{e}}$ \\
\hline
\end{tabular}

\tablefoot{
\tablefoottext{a}{\citet{Alonso-Floriano2015}.}
\tablefoottext{b}{\citet{Cifuentes2020}.}
\tablefoottext{c}{Gaia} DR3 \citep{GaiaCollaboration2023}.
\tablefoottext{d}{\citet{Reiners2018}.}
\tablefoottext{e}{\citet{Hurt2022}.}
\tablefoottext{f}{\citet{Lafarga2021}.}
\tablefoottext{g}{\citet{Diaz19}.}
\tablefoottext{h}{\citet{Hobson2018}.}
}
\end{table}

\subsection{GJ\,617A - HD147379A - HIP\,79755}

The M dwarf star GJ 617A (HD147379A, HIP\,79755, J16167+672S) is a bright star classified as M0V \citep{Alonso-Floriano2015}, and it is located at a distance of \(d = 10.767 \pm 0.002\) pc \citep{GaiaCollaboration2023}; further details are provided in Table\ref{Table1}. This star forms a common proper motion pair with a fainter companion (GJ 617B, a BY Dra variable) at a projected separation of 1.07 arcminutes, or approximately 690 AU at the system distance \citep{Stalport2023}. Rotational broadening was detected in CARMENES data \citep{Reiners2018}, which permitted measurement of the projected rotational velocity: \(v \sin i = 2.7 \pm 1.5\) km s\(^{-1}\).

In terms of stellar activity, GJ 617A exhibits weak chromospheric emission in Ca {\sc ii} H and K, with a mean $S$-index of 1.53 derived from HIRES data \citep{Butler2017}. Although \citeauthor{Butler2017} reported the absence of H$\alpha$ emission in most of their spectra, CARMENES data indicated that H$\alpha$ is observed in absorption. In contrast, \cite{Newton2017} detected H$\alpha$ emission in GJ 617A, which differs from previous findings, while \cite{Gizis2002} also recorded H$\alpha$ in absorption. This discrepancy reflects the complexity of chromospheric processes in  GJ 617A.

\citet{Reiners2018} estimated the star's rotation period to be approximately \(P \approx 31\) days based on the relationship between X-ray activity and rotational period from \citet{Reiners2014}. The uncertainty in this estimate is about ±20 days, as the X-ray values of individual stars show considerable scatter around the activity-rotation relationship.

According to \citet{Jeffers2018}, stars with rotation periods of 10 days or longer ($P_{\text{rot}} \geq 10$ days) typically lack H$\alpha$ emission, indicating low activity levels, whereas faster rotators ($P_{\text{rot}} < 10$ days) usually exhibit H$\alpha$ emission and are therefore considered active.
\citet{Hobson2018} confirmed the presence of an exoplanet previously reported using CARMENES data by \citet{Reiners2018}. This planet, GJ~617Ab, has an orbital period of 86.7 days and a minimum mass of 31.29 Earth masses, and is potentially located within the habitable zone near its inner boundary.
By combining observations from multiple facilities, \citet{Hobson2018} further refined the planetary parameters. In this joint analysis, the previously reported $\sim$500-day signal was not confirmed, indicating that it was likely spurious.

\begin{figure}[h!]
    \centering
    \includegraphics[width=\linewidth]{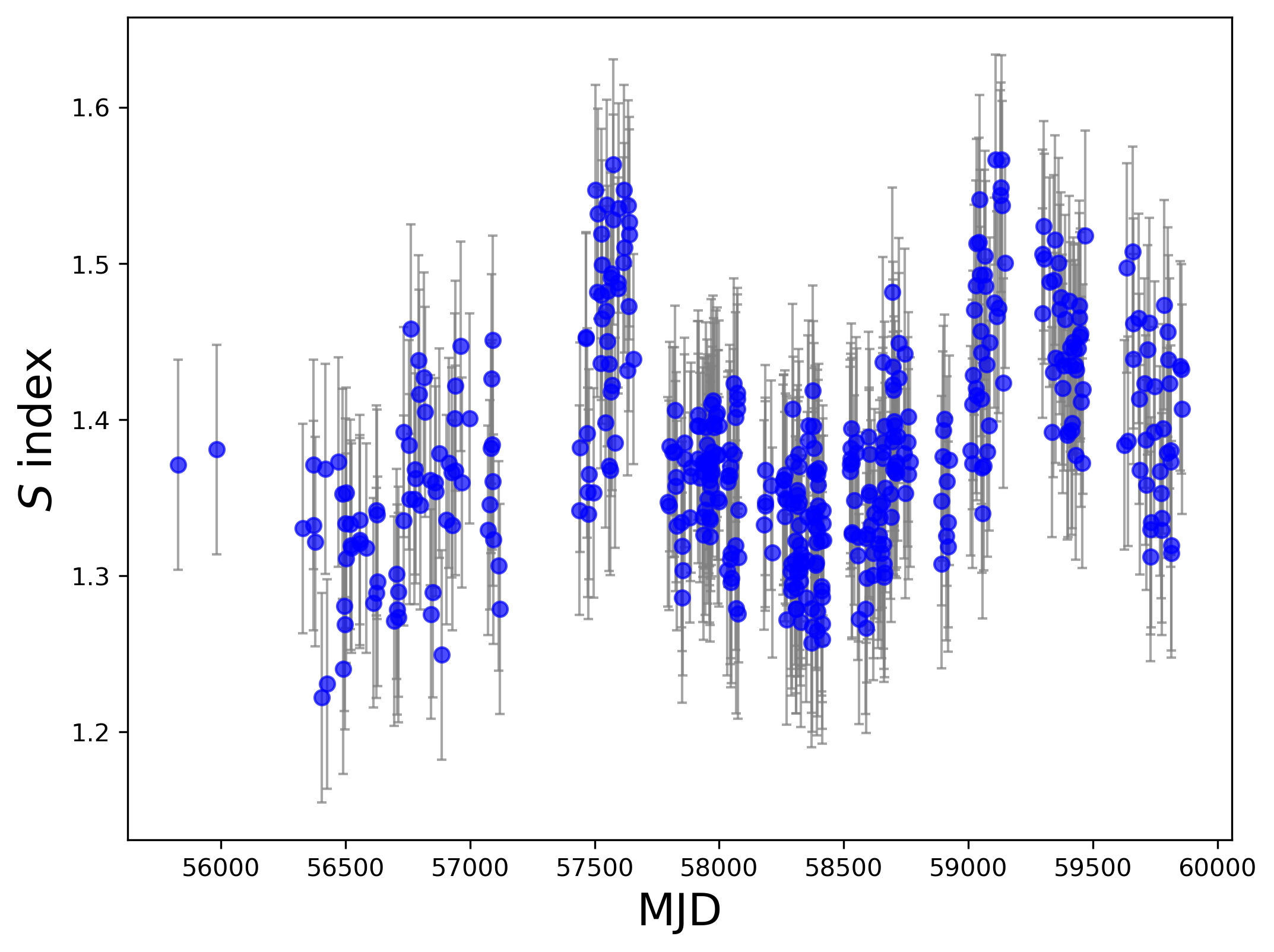} 
    \caption{Time series of the $S$-index for GJ 617A. Each point represents a SOPHIE spectrum.}
    \label{fig:serie_temporal2_GJ617}
\end{figure}

\begin{figure}[h!]
    \centering
    \includegraphics[width=\linewidth]{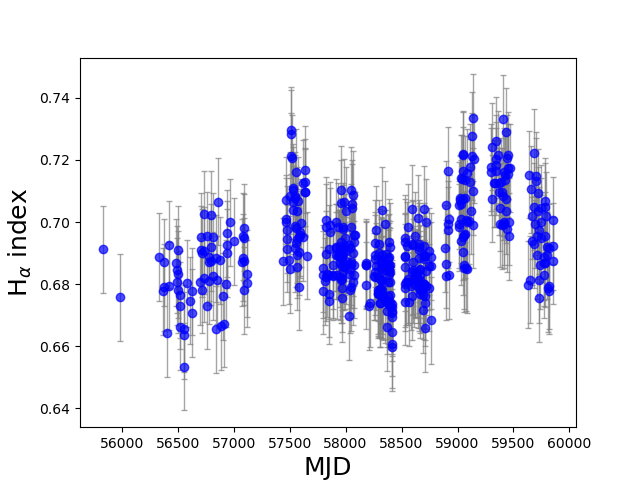} 
    \caption{Time series of the H$\alpha$ index for GJ 617A. Each point represents the spectra (data) obtained with the SOPHIE spectrograph.}
    \label{fig:serie_temporal1_GJ617}
\end{figure}

Figure \ref{fig:serie_temporal2_GJ617} illustrates the evolution of the $S$-index. The vertical axis shows the $S$-index, with a mean value of 1.27 and a standard deviation of 0.06.
Figure \ref{fig:serie_temporal1_GJ617} shows the temporal variation of the H$\alpha$ index over 10 years (2014-2024) using the same timescale as the $S$-index series. Flares should have been removed with the procedure outlined in previous section.

The two indicators, the $S$ and $H\alpha$ indices, exhibit similar variation patterns over time, suggesting a good correlation between them, with a Pearson correlation coefficient of $R = 0.75$. This is further quantified in Fig.~\ref{correlacion}, which presents the correlation between the successive differences in the H$\alpha$ and $S$-index activity indicators with $R = 0.73$.

\begin{figure}[htb!]
    \centering
    \includegraphics[width=\linewidth]{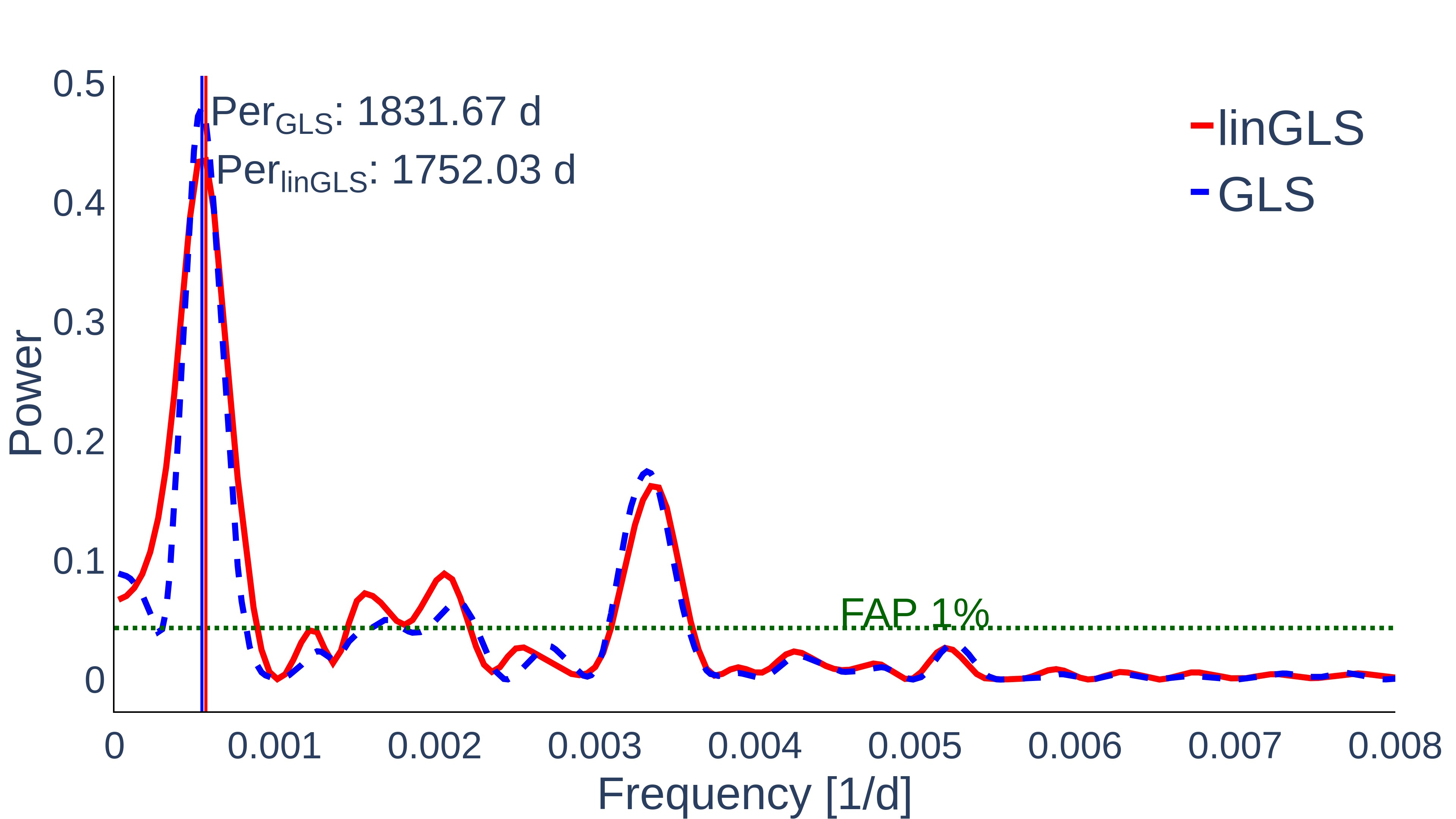} 
    \caption{Generalised Lomb--Scargle (blue) and LinGLS (red) periodograms of GJ 617A using the $S$-index. The main peaks are at (1831.67 $\pm$ 838.75) d (GLS) and (1752.03 $\pm$ 767.40) d (LinGLS).}
    \label{fig:GLS_LinGLS_GJ617_1}
\end{figure}

\begin{figure}[htb!]
    \centering
    \includegraphics[width=\linewidth]{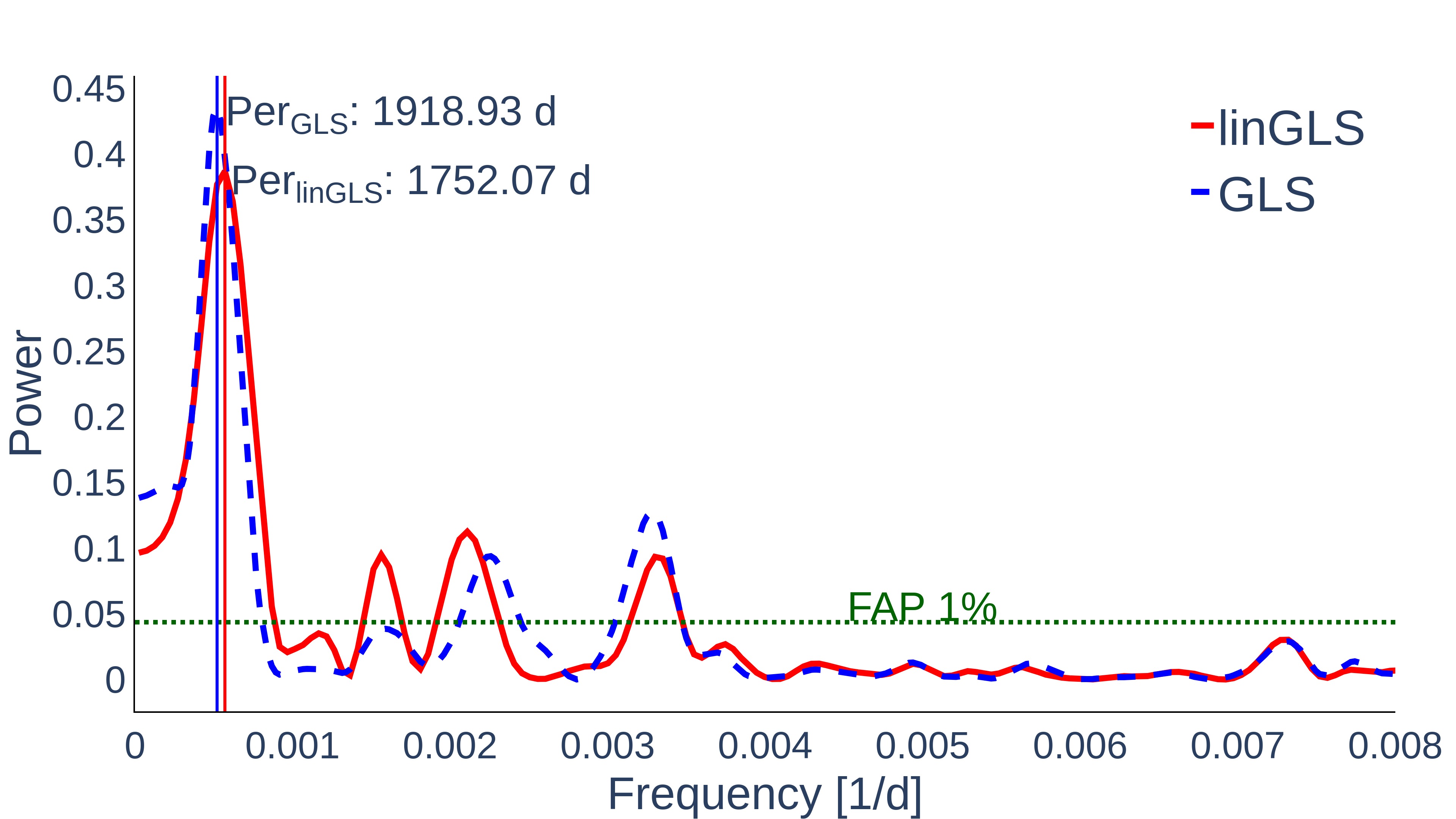} 
    \caption{Generalised Lomb--Scargle (blue) and LinGLS (red) periodograms of GJ 617A using the H$\alpha$ index. The main peaks are at (1918.93 $\pm$ 920.57) d (GLS) and (1752.07 $\pm$ 767.43) d (LinGLS).}
    \label{fig:GLS_LinGLS_GJ617_2}
\end{figure}

Subsequently, to search for long-term activity cycles, we computed the GLS and Lin-GLS periodograms. Figures \ref{fig:GLS_LinGLS_GJ617_1} and \ref{fig:GLS_LinGLS_GJ617_2} each present both periodograms superimposed corresponding to the $S$-index and the H$\alpha$ index. In the figures, prominent peaks indicate potential activity cycles. One of these peaks, corresponding to a cycle of approximately 1800 days, stands out in both models, suggesting the presence of a clear and significant cycle in both indicators (i.e. the $S$ and $H\alpha$ indices). When a long-term variation is accounted for simultaneously with the Lin-GLS model, the peaks are found at around 1700 days. All reported periods were obtained considering a FAP threshold of less than 10$^{-4}$.

\begin{figure}[htb!]
    \centering
    \includegraphics[width=\linewidth]{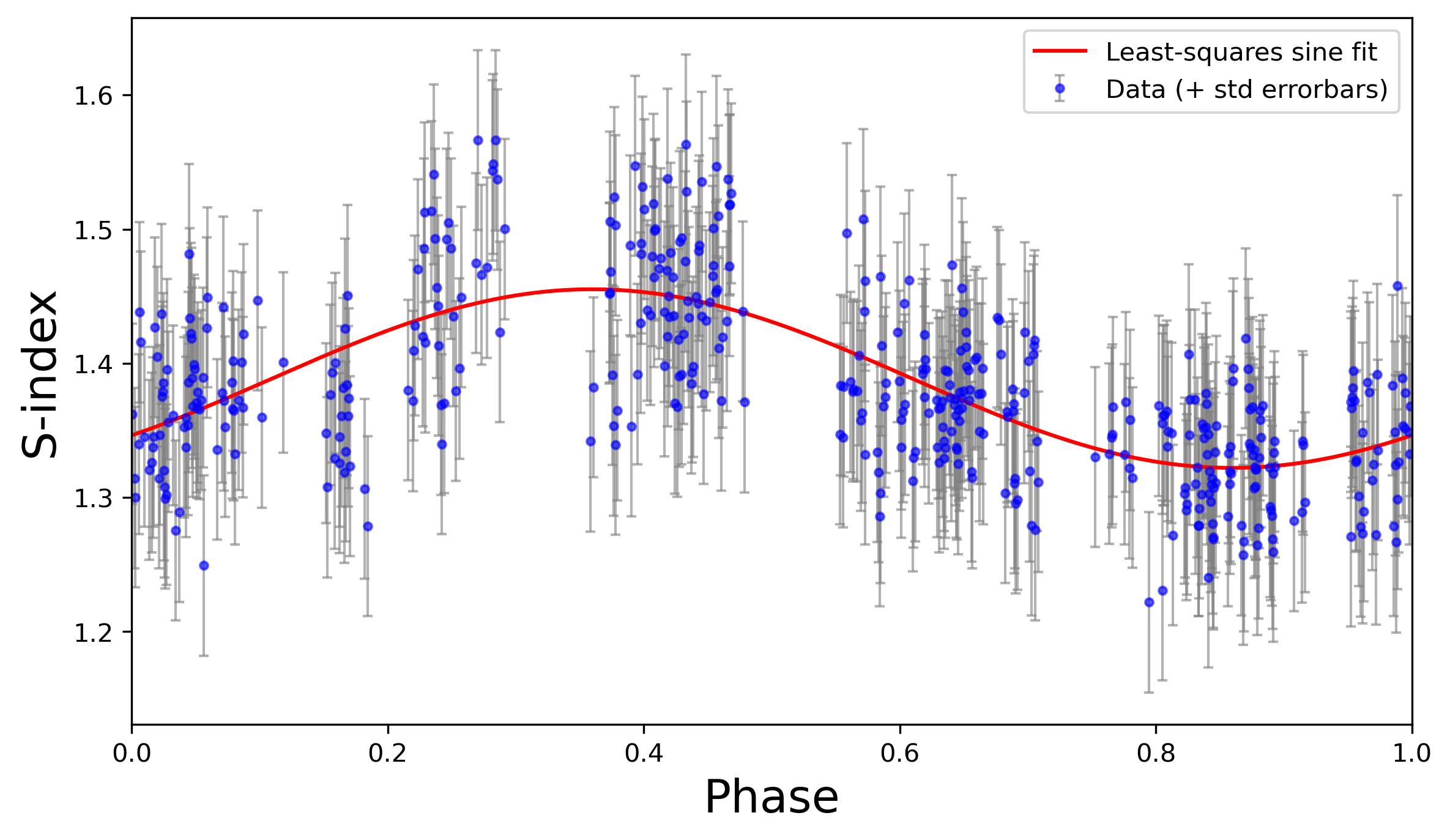} 
    \caption{Variability on GJ 617A. Phase-folded time series of the $S$ activity index versus phase, with datasets indicated by blue circles and the best-fit harmonic curve shown in red. The phase folding was performed using the period obtained from the GLS analysis.}
    \label{fig:phase-folded_s_GJ617}
\end{figure}

\begin{figure}[htb!]
    \centering
    \includegraphics[width=\linewidth]{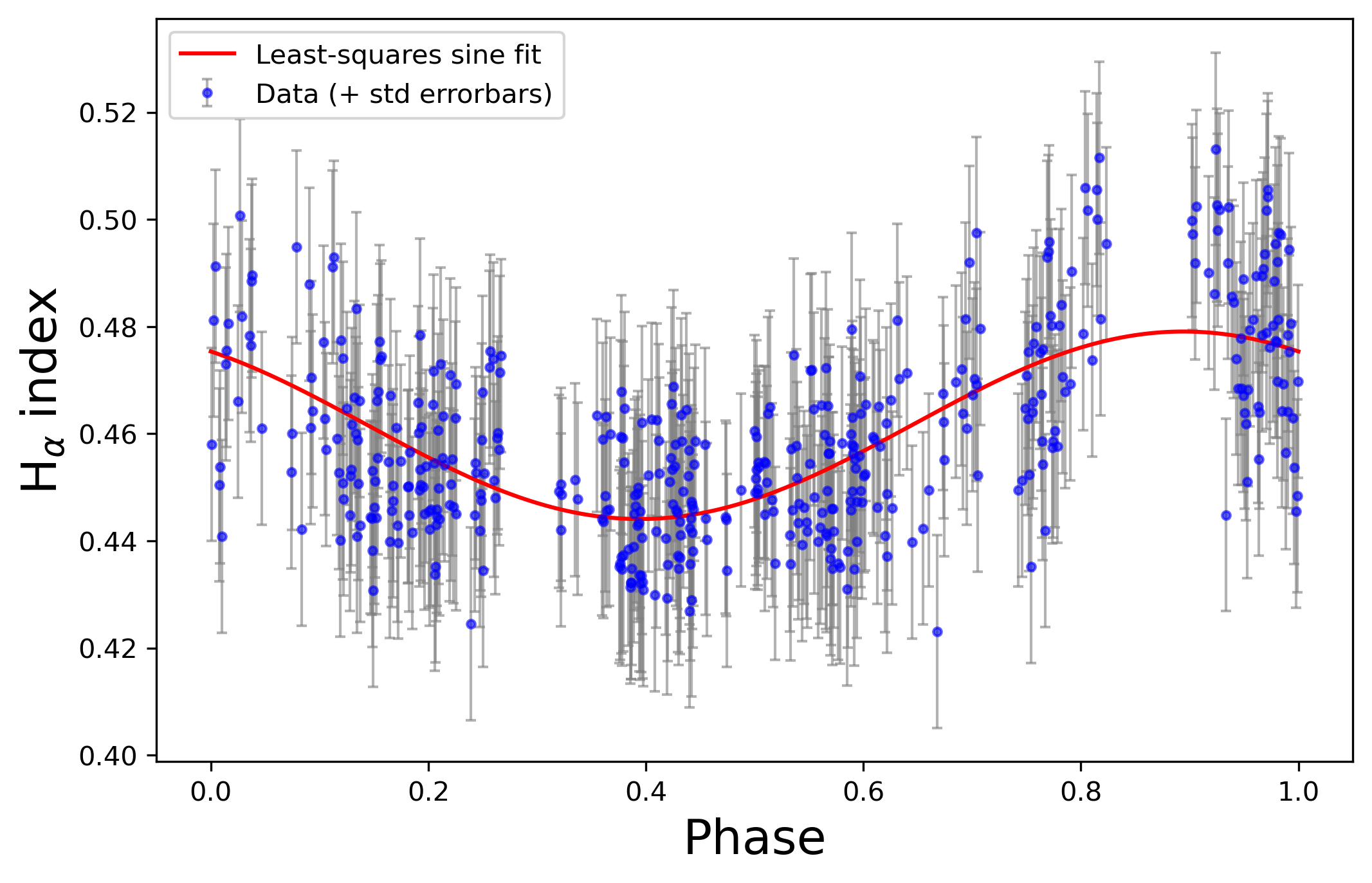} 
    \caption{Variability on GJ 617A. The phase-folded time series H$\alpha$ activity index versus the phase of the datasets is indicated with blue circles. The harmonic curve that best fits the data is in red.
     The phase folding was performed using the period obtained from the GLS analysis.}
    \label{fig:phase-folded_halfa_GJ617}
\end{figure}

\begin{figure}[h!]
    \centering
    \includegraphics[width=\linewidth]{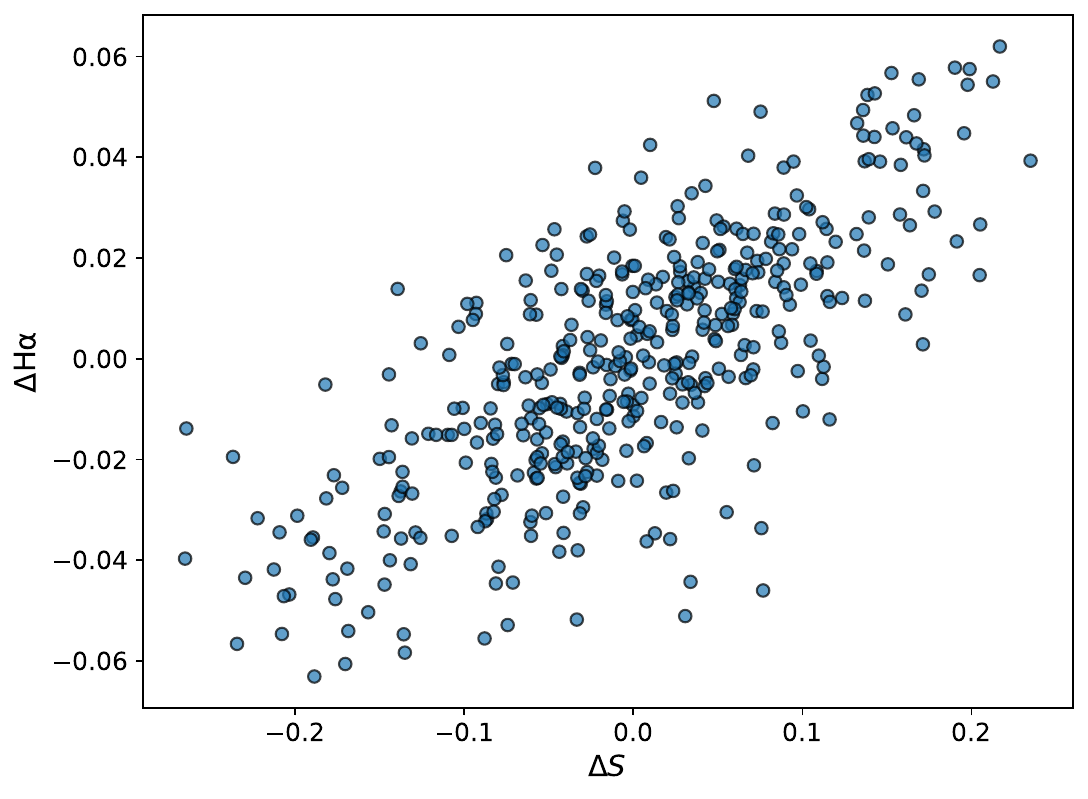} 
    \caption{Correlation between the successive differences in the H$\alpha$ and $S$-index activity indicators for GJ 617A. The Pearson correlation coefficient is R=0.73.}
    \label{correlacion}
\end{figure}

To further examine this periodicity, we constructed phase-folded time series at this period. These are shown in Fig.~\ref{fig:phase-folded_s_GJ617} regarding the $S$-index and Fig.~\ref{fig:phase-folded_halfa_GJ617} for the H$\alpha$ index, where a sinusoidal model is overplotted.
Although the phase-folded series may give the visual impression of a phase offset between the two indices, our statistical analysis indicates that they are in fact positively correlated (Pearson $R = 0.75$ for the raw indices and $R = 0.73$ for successive differences; see Fig.~\ref{correlacion}). We therefore do not interpret the apparent visual offset as evidence of anti-phase behaviour.

\subsection{GJ 411 - HD\,95735 - Lalande 21185}
The star Gliese 411, also known as Lalande 21185 or HD95735, is one of the closest M dwarfs to the Solar System. Its physical parameters are given in Table \ref{Table1}. Over the years, this star has been a primary target in the search for exoplanets. Early astrometric studies suggested the possible presence of a planetary companion, but these claims were never confirmed. More recent RV studies have provided a clearer view of the system.

\cite{Butler17} detected a periodic signal of 9.9 days, which they attributed to a planet, GJ 411b. However, \cite{Diaz19} and later \cite{Stock20} identified a more robust signal with a 12.95-day period, corresponding to a super-Earth with a minimum mass of 3.8 M$_\oplus$. \cite{Stock20} also identified a periodicity of $\sim 2900$ days, initially interpreted as a magnetic activity cycle, although \cite{Rosenthal2021} subsequently attributed it to a second planet, GJ 411c.

The study by \cite{Hurt2022} confirms the existence of the long-period planet GJ 411c. The authors also report a 200-day signal whose nature could not clearly be established. This candidate signal has yet to be definitively confirmed, but it shows promising signs of being a genuine exoplanet.

\begin{figure}[htb!]
    \centering
    \includegraphics[width=\linewidth]{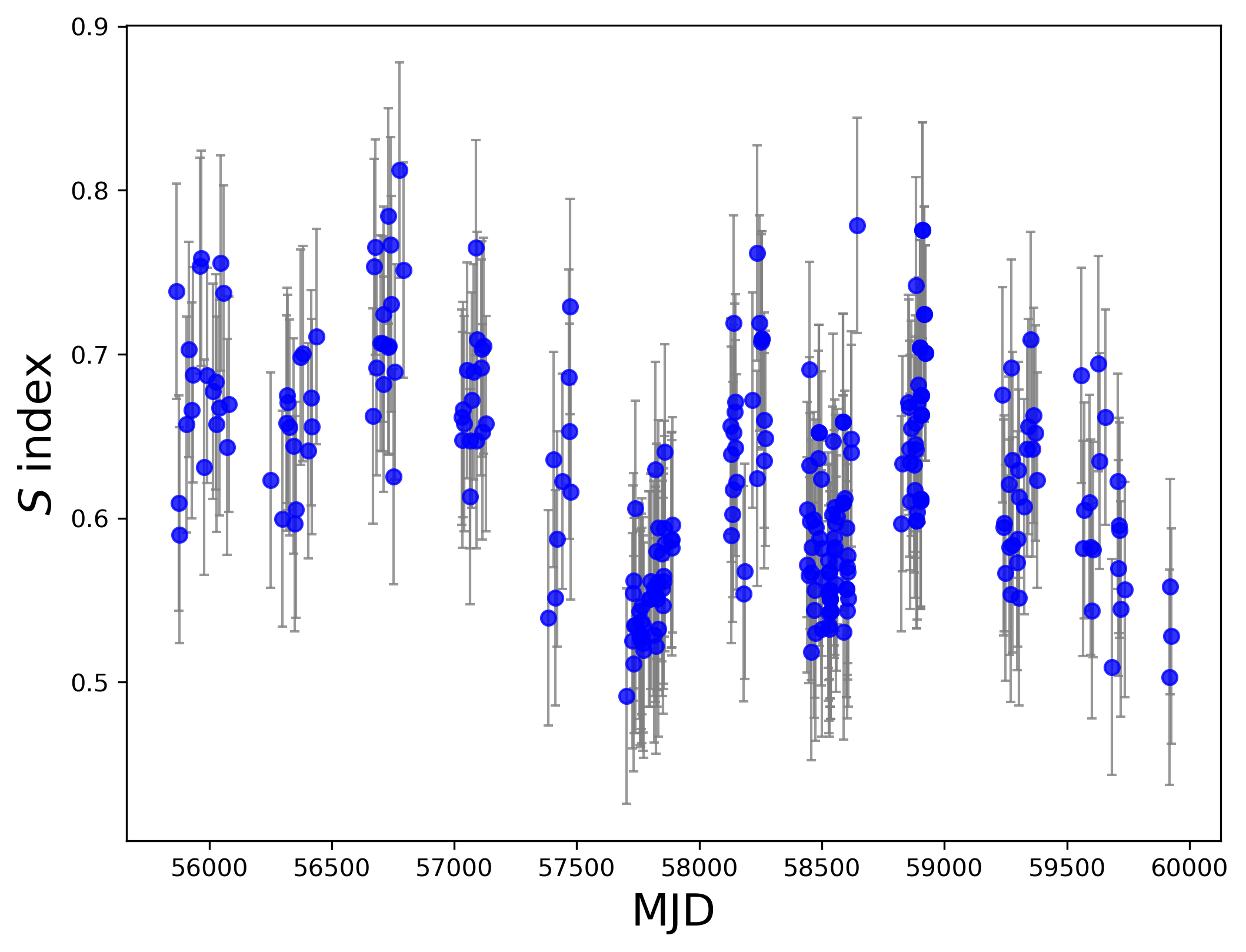} 
    \caption{Time series of the $S$-index for GJ 411. Each point represents the spectra (data) obtained with the SOPHIE spectrograph.}
    \label{fig:serie_temporal1_GJ411}
\end{figure}

\begin{figure}[htb!]
    \centering
    \includegraphics[width=\linewidth]{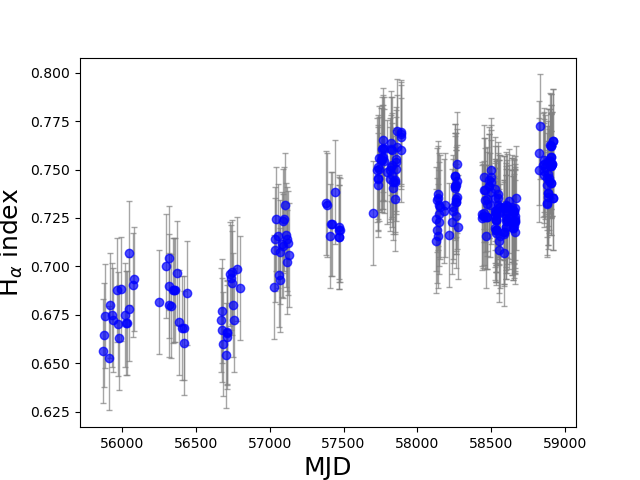} 
    \caption{Time series of the H$\alpha$ index for GJ 411. Each point represents the spectra (data) obtained with the SOPHIE spectrograph.}
    \label{fig:serie_temporal2_GJ411}
\end{figure}

This work is based on 284 SOPHIE spectra acquired between 2008 and 2024. Figure \ref{fig:serie_temporal1_GJ411} shows the time series for the $S$-index of GJ 411, which has a mean value of \(\langle S \rangle = 0.65 \) and a standard deviation of 0.09. We show the time series for the H$\alpha$ index in Fig. \ref{fig:serie_temporal2_GJ411}. We observed that the temporal behaviour of the $S$-index and H$\alpha$ index do not exhibit a correlation over time, indicating distinct patterns in their respective time series.

\begin{figure}[htb!]
    \centering
    \includegraphics[width=\linewidth]{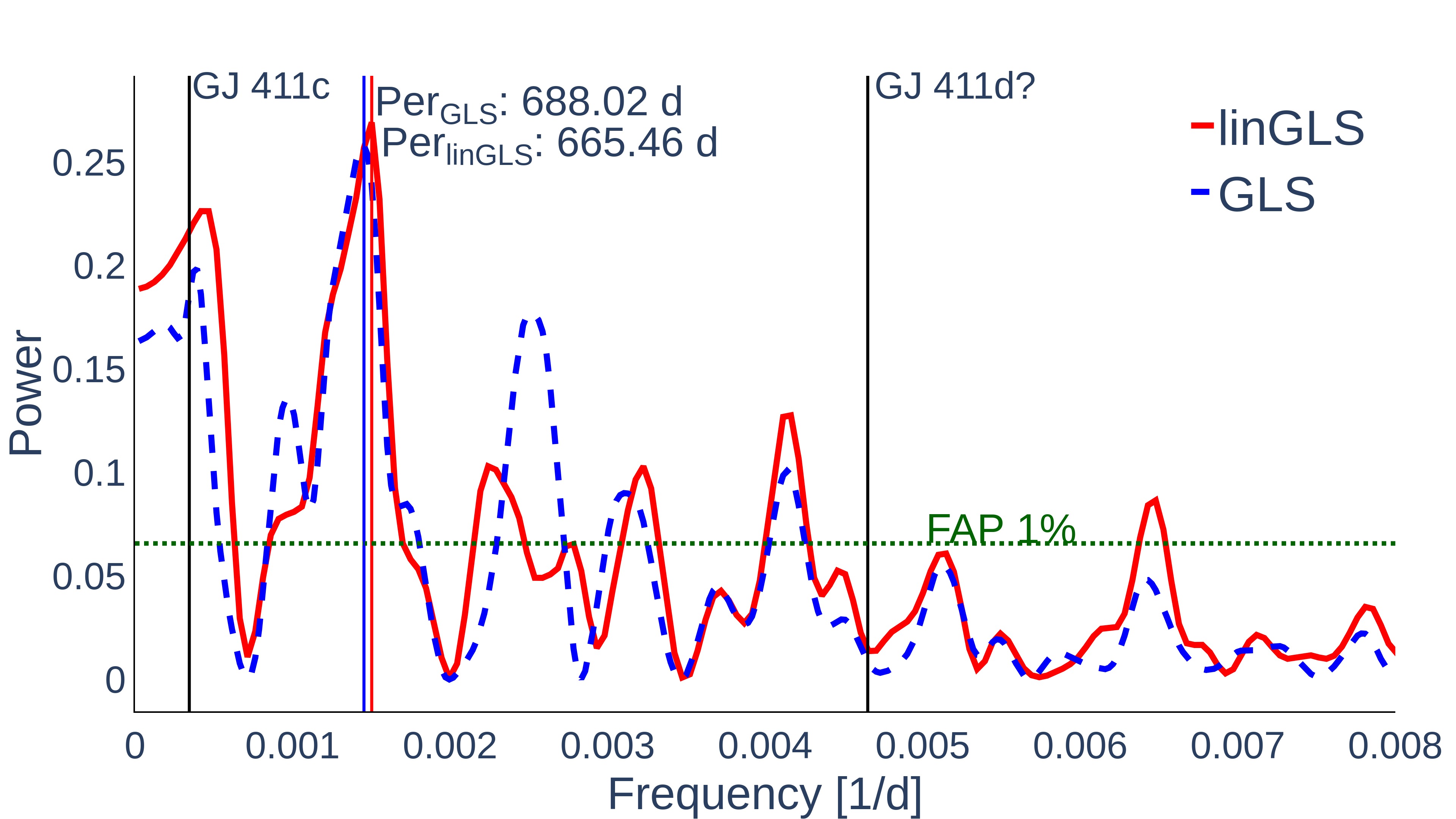} 
    \caption{Generalised Lomb--Scargle (blue) and LinGLS (red) periodograms of GJ 411 using the $S$-index. The main peaks are at (688.02 $\pm$ 118.34) d (GLS) and (665.46 $\pm$ 110.70) d (LinGLS). The vertical black lines mark the orbital period of GJ 411c ($\sim 2900$) and an additional signal around 200 d.}
\label{fig:GLS_LinGLS_GJ411_1}
\end{figure}

Following a similar methodology for GJ 617A, we built the corresponding periodograms for both activity indicators. In Fig. \ref{fig:GLS_LinGLS_GJ411_1} we show the resulting periodograms for the $S$-index. A significant peak was found at 688.02 days using the GLS method and at 665.46 days using the LinGLS method. 

\begin{figure}[htb!]
    \centering
    \includegraphics[width=\linewidth]{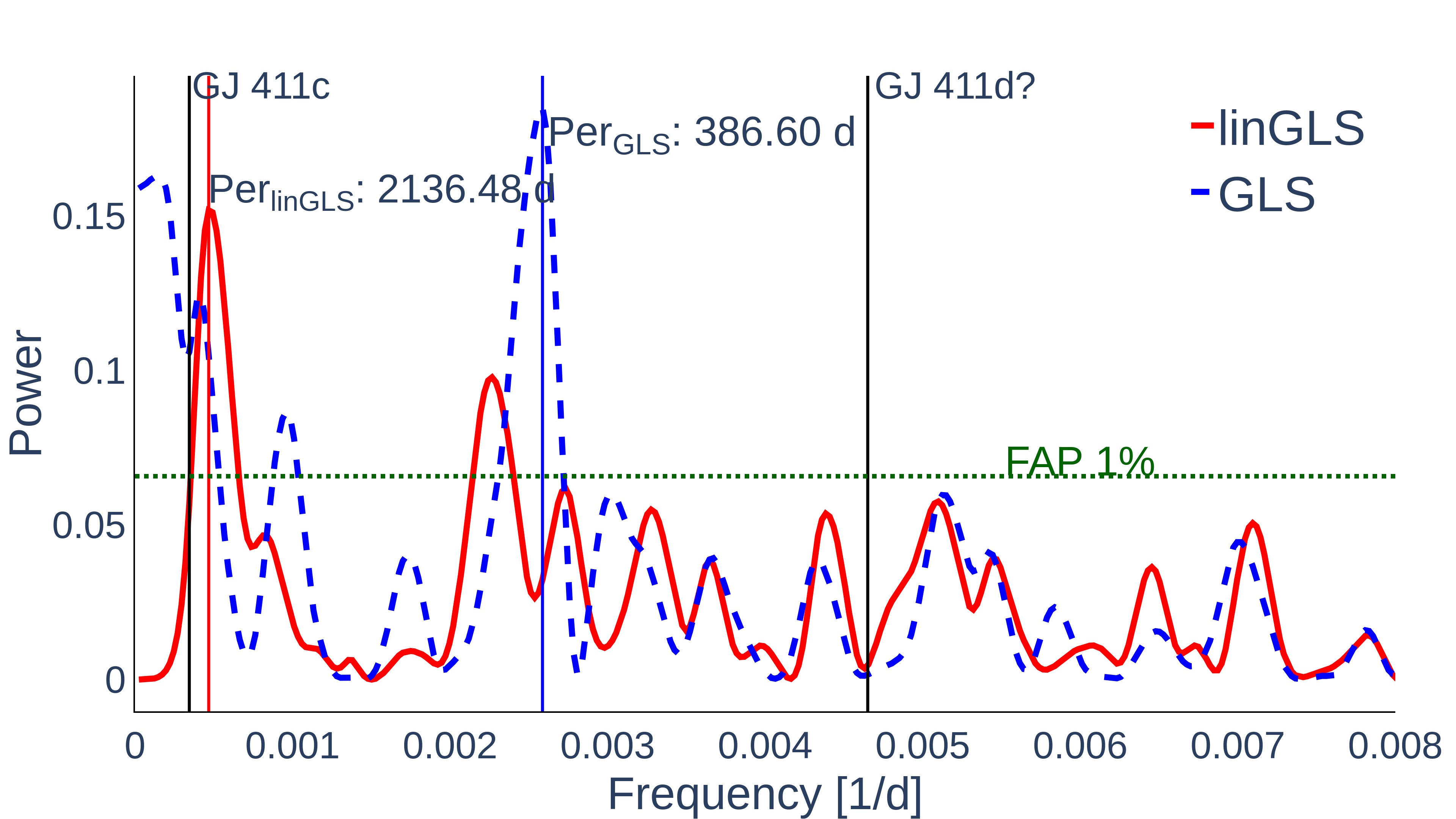} 
    \caption{Generalised Lomb--Scargle (blue) and LinGLS (red) periodograms of GJ 411 using the $S$-index after subtracting a harmonic of the detected period. The main peaks are at (386.6 $\pm$ 37.36) d (GLS) and (2136.48 $\pm$ 1141.13) d (LinGLS). The vertical black lines mark the orbital period of GJ 411c (~$\sim 2900$ d) and an additional signal around 200 d.}
    \label{fig:GLS_LinGLS_GJ411_2}
\end{figure}

To explore the low-frequency domain where \cite{Stock20} announced the additional companion, we subtracted a harmonic function of the period detected from each time series. This allowed for the identification of a second significant peak: 386 days in the GLS periodogram and 2136 days in the LinGLS periodogram (as shown in Fig. \ref{fig:GLS_LinGLS_GJ411_2}).

\begin{figure}[htb!]
    \centering
    \includegraphics[width=\linewidth]{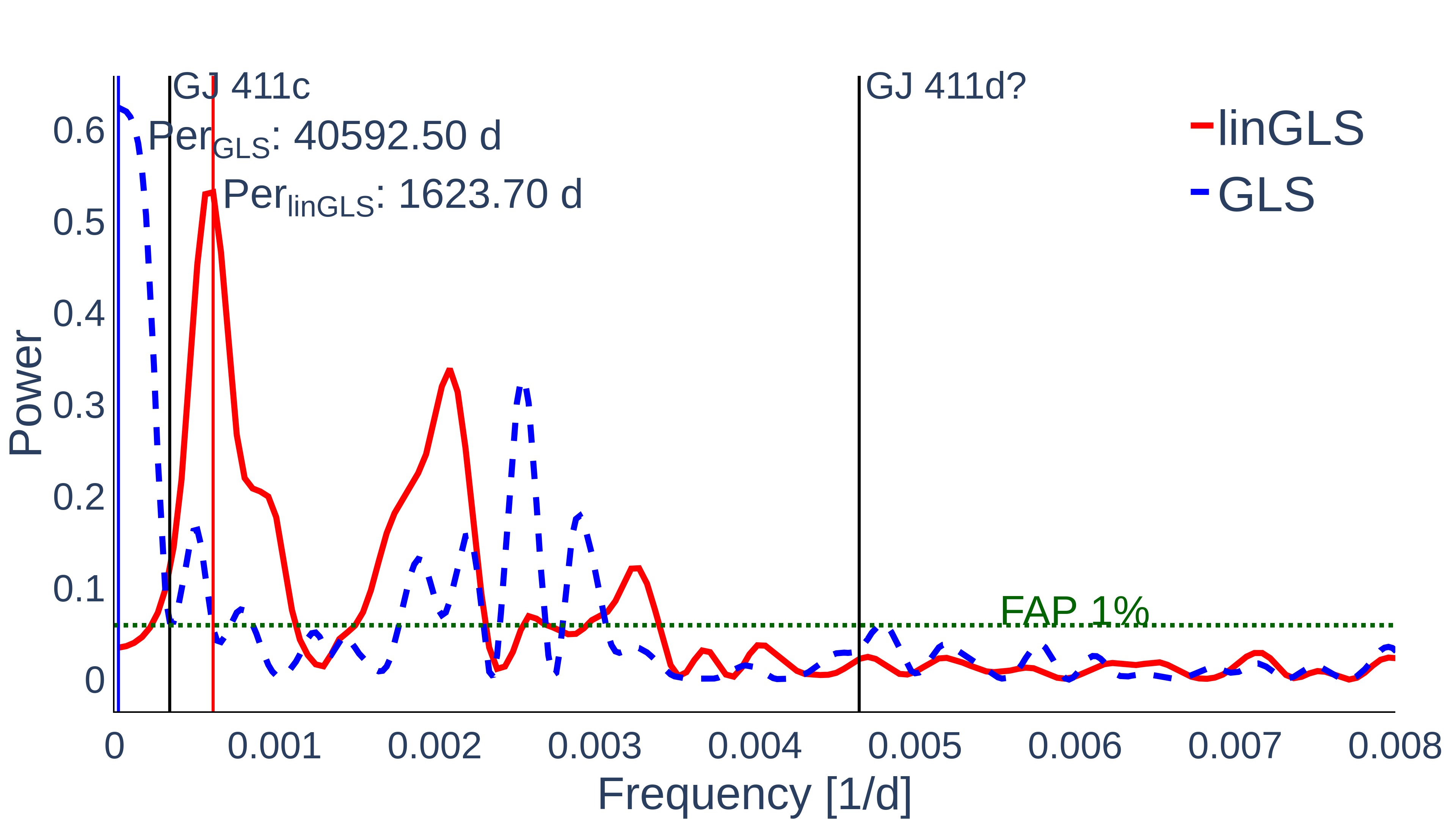} 
    \caption{Generalised Lomb--Scargle (blue) and LinGLS (red) periodograms of GJ 411 using the H$\alpha$ index. The main peaks are at (40592.50 $\pm$ 411.94) d (GLS) and (1623.7 $\pm$ 659.10) d (LinGLS). The vertical black lines mark the orbital period of GJ 411c (~$\sim 2900$ d) and an additional signal around 200 d.}
    
    \label{fig:GLS_LinGLS_GJ411_h}
\end{figure}

This analysis reveals that the quadratic polynomial term is statistically significant only in the case of GJ 411, while it is not required to describe the variability of GJ 617A. This result provides a quantitative explanation for the differences observed between the standard GLS and LinGLS periodograms. The activity indicators of GJ 411 exhibit long-term variations that cannot be characterised by a purely sinusoidal model. Such behaviour points to complex non-sinusoidal variability or secular trends consistent with a long-period magnetic cycle in evolution or a transition towards a broad minimum or due to a Waldmeier effect, which has been observed in other cool stars \citep{Robertson2013,Buccino2020,Flores2024}. Consequently, long-term RV and activity monitoring over decadal timescales are required to characterise the nature of this stellar activity.

We also performed a periodogram analysis for the H$\alpha$ index (see Fig. \ref{fig:GLS_LinGLS_GJ411_h}). A low-frequency rise was observed, as expected from the appearance of the time series. In the Ca \ii\- index, the dominant signal appears at 2136 days, whereas in the H$\alpha$ index, the most significant period is approximately 1624 days. The latter suggests the presence of a long-term activity cycle that is not consistently detected in both indices. These differences highlight that the dominant periodicities inferred from different activity indicators may vary, as also reported by \citep{Mignon2023} for several M dwarfs.

\subsection{Stellar population membership for GJ 411 and GJ 617A}

Using proper motions and parallaxes from Gaia DR3 \citep{GaiaCollaboration2018}, we derived the Galactic space-velocity UVW components for both stars by following the procedure detailed in \citet{Jofre2015}. We find that $(U, V, W) = (-56.08, -48.56, -67.10)$ km s$^{-1}$ for GJ 411 and $(U, V, W) = (-0.35, -24.95, 11.46)$ km s$^{-1}$ for GJ 617A. 

Then, based on these space-velocity components and using the membership formulation of \citet{Reddy2006}, we determined that the probability of belonging to the thin-disc population is $\sim$98\% for GJ 617A. The UVW values for GJ 411 are consistent with thick-disc membership, with a probability of $\sim$93\%. 

\section{Searching for hints of short-term photometric activity in TESS data}

The Transiting Exoplanet Survey Satellite \citep[][]{ricker2015} is a NASA space mission launched in 2018 whose main purpose is to look for and characterise transiting planets around bright and nearby stars. However, thanks to its high-cadence observing modes, TESS observations are also valuable for detecting stellar variability, including pulsations, rotational modulation, and flares.
Therefore, as a complement to the long-term spectroscopic study, we used photometric datasets from TESS to characterise the short-term variability of the two stars. Furthermore, simultaneous photometry and activity indices help determine whether stellar activity is dominated by spots with cooler surface features, which make stars redder when fainter, or by plages, which make stars bluer with higher activity \citep{Hall2009}

\subsection{Rotation period}

The stars GJ 411 and GJ 617A were observed by TESS in several sectors with multiple cadences. Table \ref{tab_tessinfo} provides a summary of these observations. For both stars, we used the tools available in the \textsc{Lightkurve} Python package \citep{lightkurve2018} to analyse the light curves resulting from the pre-search data conditioning simple aperture photometry (PDCSAP) data processed with the TESS Science Processing Operations Center (SPOC) pipeline \citep{jenkins2016}. After removing bad points, we ran a GLS periodogram to search for signs of sinusoidal modulation. The rotation period, $\mathrm{P_{rot}}$, was determined as the inverse of the frequency with the highest peak found by GLS. To ensure that $\mathrm{P_{rot}}$ was reliable, we required it to satisfy the following conditions:
\begin{enumerate}
  \item False alarm probability (FAP) $\leq  0.01$.
  \item The standard deviation of the residuals after removing the periodic signal is smaller than or, at most, equal to the light curve's standard deviation before removing the sinusoidal modulation.
  \item More than 50$\%$ of the available cadences show the period value identified by GLS.
  \item After a careful by-eye inspection, variability is visible in the phase-folded light curve.
  \item The value of the period found differs from the duration of a sector, the duration of half a sector, and the duration of the full light curve.
\end{enumerate}

For GJ 411, before eliminating bad points, it was necessary to apply a median filter to the photometric data to eliminate the still present systematics. For this target, none of the detected signals turned out to be significant, and hence there is no evidence of rotational variability in the TESS data analysed.
The absence of a clear detection is consistent with the ongoing uncertainty in the literature surrounding the true rotation period of GJ 411. For instance, \cite{Diaz19} proposed a rotation period of 56 days based on photometric variations. However, more recent analyses using SPIRou spectropolarimetric data have yielded unusual results: \cite{Donati2023} report a strange and poorly understood modulation with a period of 427 days, while \cite{Fouque2023} propose a very long rotation period of 478 days based on variations in the longitudinal magnetic field (Bl).

\begin{figure}[htbp] 
\centering
\includegraphics[width=1.05\columnwidth]{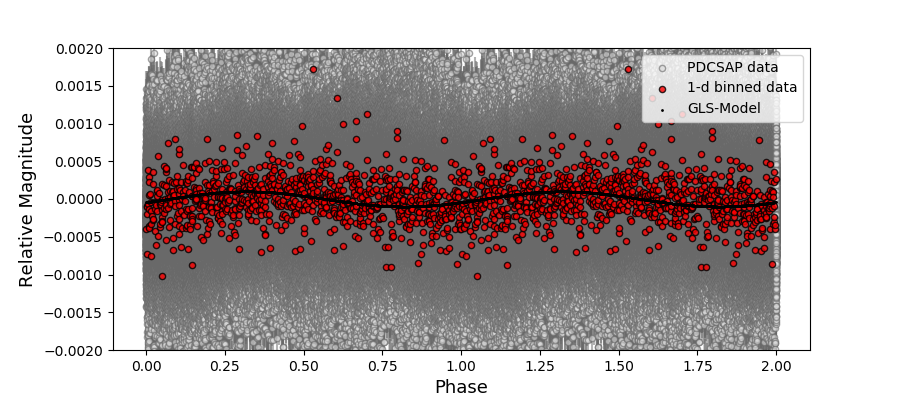}
\includegraphics[width=1.05\columnwidth]{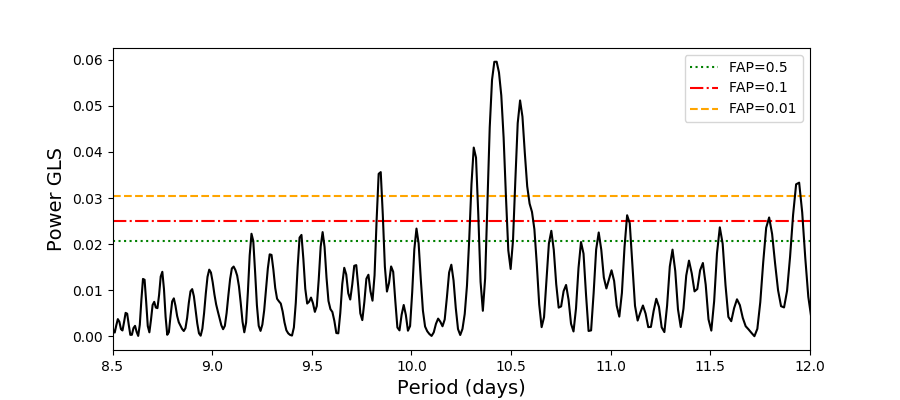}
\caption{TESS photometric rotation signature of GJ 617A. Top panel: Zoom-in view of the phase-folded 2-min cadence TESS light curve of GJ 617A considering a period of 10.4 days. Grey and red symbols indicate the 2-min cadence and 1-day binned TESS data, respectively. The solid line represents the best sinusoidal fit to the data. Bottom panel: GLS periodogram of the 1-day binned 2-min cadence TESS light curve. Horizontal lines indicate different FAP values.}
\label{FigPhoto}
\end{figure}

\begin{figure}[htbp] 
\centering
\includegraphics[width=0.5\columnwidth]{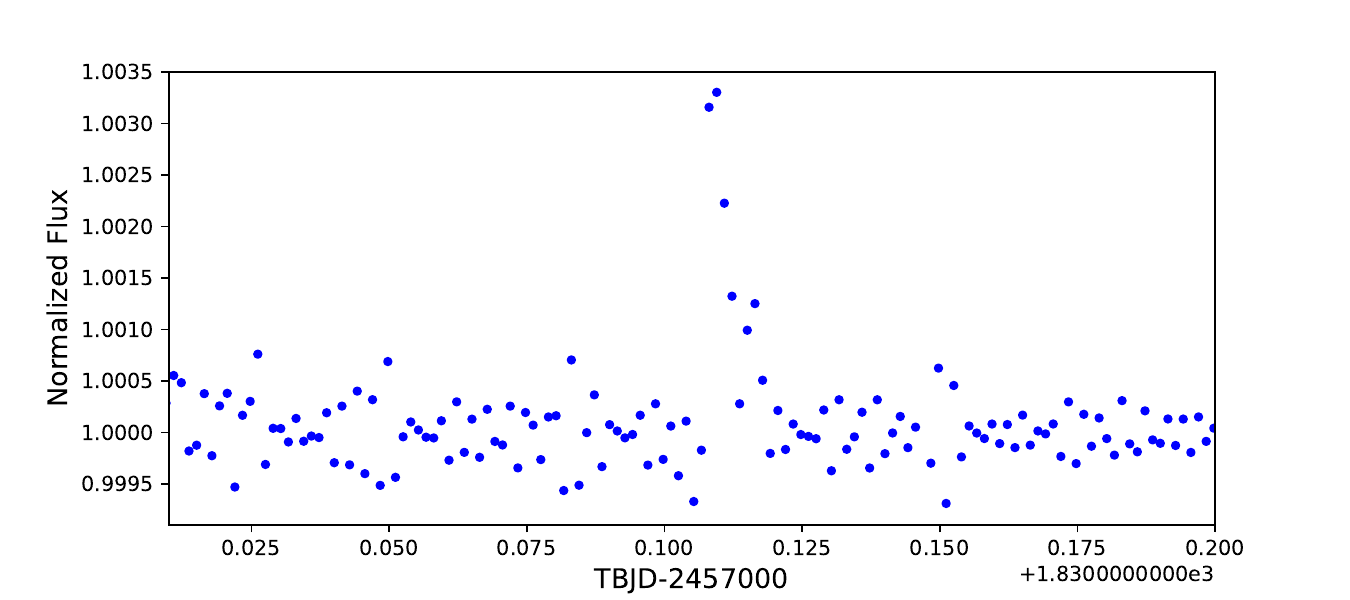}
\includegraphics[width=0.5\columnwidth]{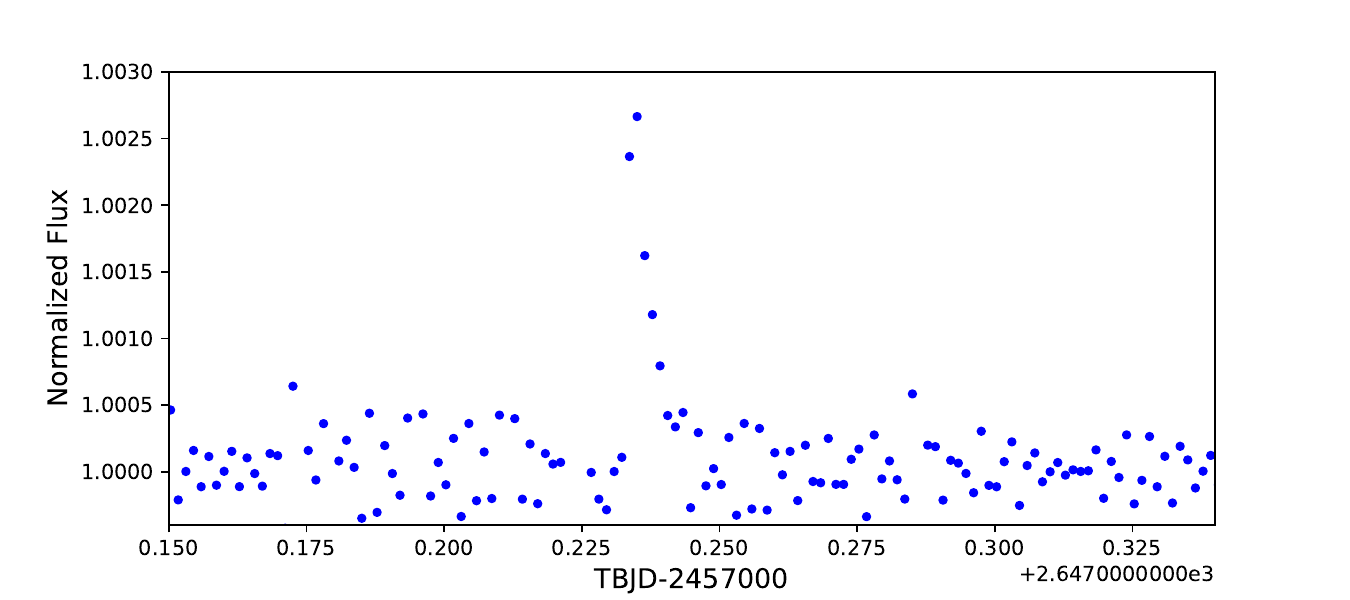}
\includegraphics[width=0.5\columnwidth]{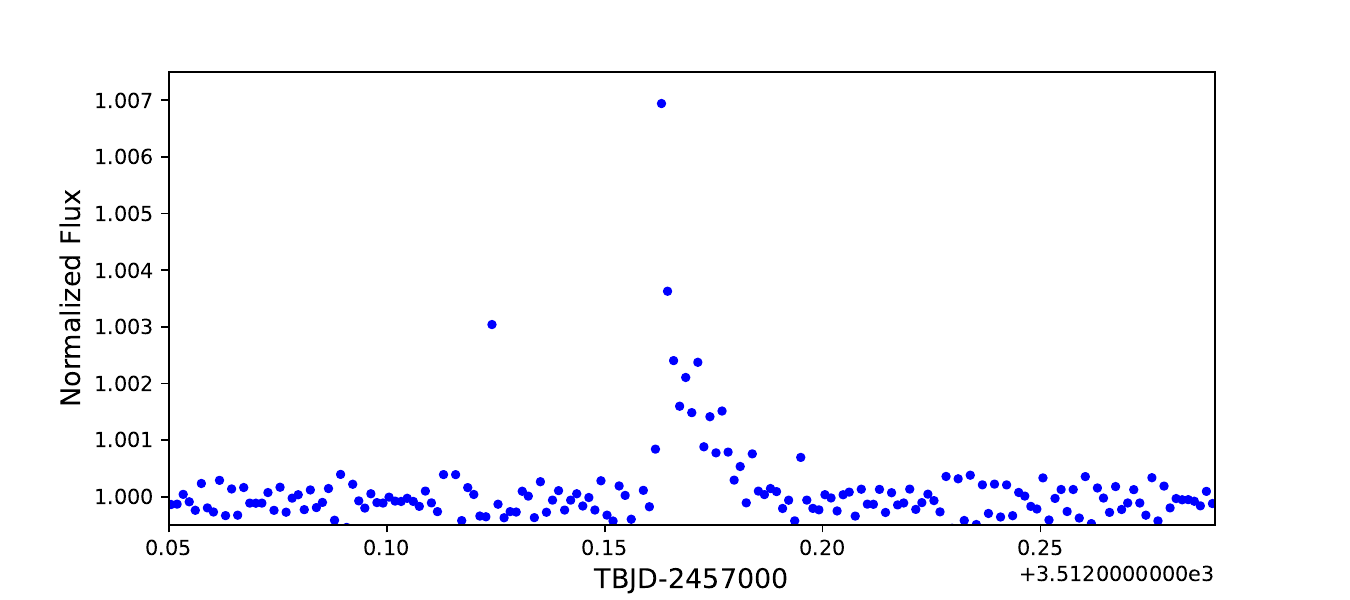}
\includegraphics[width=0.5\columnwidth]{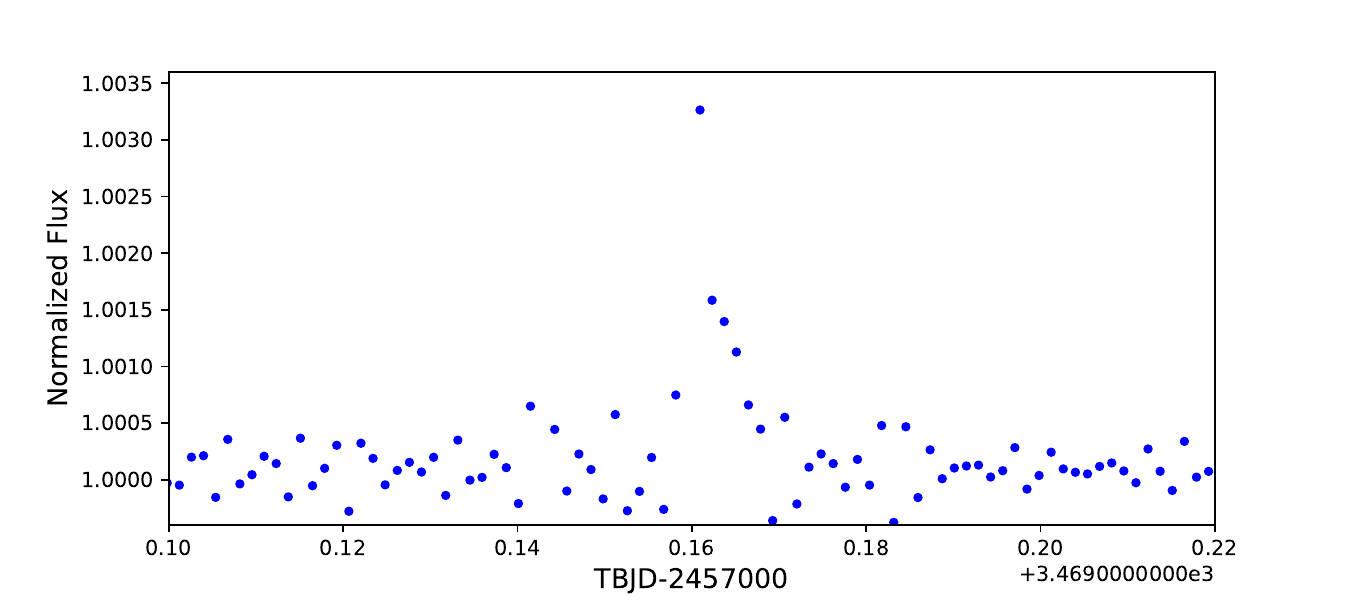}
\caption{Examples of flare events found in GJ 617A.}
\label{flares}
\end{figure}

In contrast, for GJ 617A, we found a reliable rotation period of 10.412 $\pm$ 0.055 days with a photometric amplitude of Amp = 0.0001 mag in the 2-min cadence data considering 38 TESS sectors. Inside errors, $\mathrm{P_{rot}}$ values were also found after running the GLS periodogram on the 200s, 1800s, and 20s cadence TESS light curves. The 600s cadence was the only one that showed a different peak value in the periodogram. The $\mathrm{P_{rot}}$ value around 10 days detected in our analysis agrees with the first harmonic found in previous works using spectroscopic data \citep{Hobson2018, Reiners2018} and 28 sectors of TESS photometry \citep{Stalport2023}. In this study, we did not recover the $\sim$21-day signal found by \cite{Stalport2023}. One possible reason is that in this work, we used PDCSAP data already corrected by systematics, while in \cite{Stalport2023}, the authors employed simple aperture photometry data and performed their own de-trending to eliminate the systematics. In Fig. \ref{FigPhoto} we present the phase-folded 2-min cadence light curve considering a period of 10.4 days and the corresponding GLS periodogram.

Finally, we explored whether the rotation period values measured in each individual TESS sector evolve along with the minimum and maximum values of the $S$ and H$\alpha$ indices. After running a GLS periodogram on the data and a thorough visual inspection, no correlation was found between the short- and long-term variability.
\begin{table}
\centering 
\caption{Summary of TESS observations.}
\label{tab_tessinfo}  
\begin{tabular}{cccc}
\hline
\hline
\noalign{\smallskip}  
Star name & {TIC ID} & Cadence  &  Sector \\ 
          &          &   (s)    &         \\
\hline
\noalign{\smallskip}           
GJ 411 & 353969903 & 120 & 22, 48 \\
      &           & 600 &  48 \\
      &           & 1800 &  22 \\
\hline
\noalign{\smallskip}
GJ 617A & 230073581 & 20 & 40, 41, 47-60,\\
        &           &    &  73-82\\  
        &           & 120 &  14-21, 23-26,\\ 
        &           &    &  40, 41, 47-60,\\   
        &           &    &  73-82 \\
        &           & 200 &  56-60 \\      
        &           & 600 &  40, 41, 47-55\\         
        &           & 1800 &  14-21, 23-26\\     
\hline 
\end{tabular}
\end{table}

\subsection{Flares}
\label{sec:flares}

To look for flaring events in the stars, we followed the methodology implemented in previous works \citep{petrucci2024, martioli2024}. First, a Savitzky-Golay filter with an optimum value of window length (wl = 1) was applied to the TESS light curve to remove systematics and intrinsic stellar variability. Then, the \textsc{AltaiPony} code \citep{davenport2016, ilin2021} was used to identify flare candidates. We repeated the same process in all the available sectors and cadences. Bona fide flares were then confirmed after a careful by-eye inspection, and their bolometric energy, E$_{\rm bol}$, was computed as indicated in Eq. (1) of \cite{petrucci2024}. As an additional check, we examined if these bona fide events were detected in the same sector at different cadences.

We found no signs of flaring activity in the TESS light curve of GJ 411. For GJ 617A, we identified nine flare events with bolometric energies ranging from 3.44 $\times$ 10$^{33}$ erg s$^{-1}$ to 4.93 $\times$ 10$^{32}$ erg s$^{-1}$.

\section{Summary and conclusions}

M dwarfs play a crucial role in exoplanet research, as their small sizes and low luminosities make them particularly favourable for the detection of Earth-sized planets within their HZs. Nonetheless, their intrinsic stellar activity—manifested through phenomena such as flares, rotational modulation, and long-term activity cycles—introduces significant challenges for both the detection of planetary signals and the characterisation of exoplanetary atmospheres.

With this in mind, in the present work, we characterised the stellar activity of two early M stars with planets, GJ 617A and GJ 411, based on 13 years of SOPHIE data. For GJ 617A, we built an $S$ and H$\alpha$ index time series. From analysis with different periodograms, we reported for the first time a $\sim$ 1800-day periodic signal. The observed periodicity in both indices suggests a coherent  magnetic cycle that influences multiple atmospheric layers. The positive correlation between both incremental indices, \( R = 0.75 \), is consistent with the distribution of the correlation coefficients with activity observed in Fig. 6 in \cite{Meunier2024}.

Although GJ 411 is a very inactive star, the nature of the long-term periodicities observed in its RV variations—whether of planetary origin or linked to stellar activity—remains a topic of debate \citep{Stock20, Rosenthal2021, Hurt2022}. Here, we analysed simultaneous observations of  H$\alpha$ and calcium indices  between 2011 and 2024. Our analysis revealed a primary periodicity of 688 days and a secondary long-term variability of 2136 and 1623 days in both indices, respectively. We are not certain about the origin of the 688-day signal. Because the signal is seen on a single activity index, we tentatively attribute it to a combination of a long-term variation aliasing with the window function. Furthermore, the current understanding of activity cycles disfavours cycles with such a short periodicity on low-activity-level stars. This finding underscores the complexity of the underlying dynamo processes in GJ 411. 

The detected long-term activity periods are summarised in Table~\ref{table:cycles}. From the mean $S$-index of the SOPHIE dataset for each star, we computed $\log R'_{\mathrm{HK}}$ values using the calibration of \citet{Boisse2010} for the $V-K$ colour together with the corresponding coefficients from \citet{AstudilloDefru2017} (see Table~\ref{table:cycles}).

Based on their kinematic properties, we classified GJ~617A as a thin-disc star and GJ~411 as a member of the Galactic thick disc. The difference in activity regimes can be partially explained by the rotational evolution of M dwarfs. As stars age, they lose angular momentum through magnetic braking, leading to progressively slower rotation. This reduced rotation rate decreases the efficiency of the dynamo mechanism and thus lowers the observable level of magnetic activity. Therefore, the very low level of stellar activity observed in GJ~411—evidenced by the absence of flares and rotational modulation in photometric data, as well as weak variability in the chromospheric H$\alpha$ and Ca II H and K indices—is consistent with the old age associated with the thick-disc population ($\gtrsim$ 10–11 Gyr; \citealt{Bensby2005, Reddy2006, Adibekyan2011}).

To complement this long-term study, we performed a short-term analysis using high-cadence \textit{TESS} photometry (see Table~2).
For GJ~411, we did not detect any appreciable short-term photometric variability and therefore do not confirm the rotation period of 56~days reported in previous studies.

In contrast, the \textit{TESS} photometry of GJ~617A reveals a sinusoidal modulation with a period of 10.412~days, which we associate with the first harmonic of its rotation period of 22~days.
Additionally, we identify several stochastic events attributable to flares, with energies ranging from $1 \times 10^{33}\,\mathrm{erg\,s^{-1}}$ to $4.93 \times 10^{32}\,\mathrm{erg\,s^{-1}}$.

None of the detected periods for these stars coincide with the previously reported planetary signals. This suggests that the variations observed in the activity indicators are associated with magnetic variability driven by stellar dynamo processes, thus strengthening the validity of the published planetary detections.

\begin{figure}[htb!]
    \centering
    \includegraphics[width=0.7\linewidth]{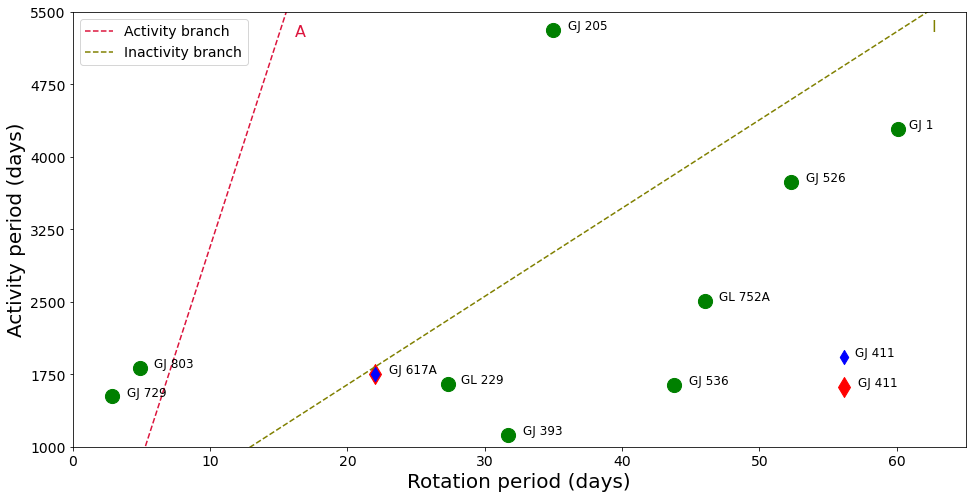}
    \caption{Activity period (days) versus rotation period (days). The dashed red lines represent the active branch (A), while the dashed olive lines represent the inactive branch (I). These lines were taken from the work of \citet{Metcalfe2016} and were computed for solar-type stars. The green dots correspond to some M-type stars. Blue diamonds represent the results obtained in this work based on the $S$-index, and red diamonds correspond to results obtained based on the H$\alpha$ index.}
    \label{fig:branches}
\end{figure}

In order to analyse these long-term cycles in the stellar context, we placed our results in the empirical diagram $P_{cyc}-P_{rot}$ built for solar-type stars in \cite{Metcalfe2016}, see Fig. \ref{fig:branches}. Furthermore, we included the activity cycles of early and mid-M dwarfs also reported in the literature \citep{Buccino2011, IbanezBustos2020,IbanezBustos2025}. In Fig. \ref{fig:branches}, it is remarkable that GJ 617A falls in the branch associated with inactive solar-type stars. This may give the idea that rotation drives the magnetic dynamo in GJ 617A. On the other hand, the cycles detected for GJ\,411 do not fall close to any branch in the diagram. This behaviour seems to be similar to other early M star slow-rotators such as GJ 229 or GJ 536.

In summary, we have presented a long-term analysis of stellar activity based on a homogeneous dataset for two early M-type dwarf stars: GJ 617A and GJ 411. Although the stellar interiors of the stars exhibit similar physical characteristics and structure, the stars differ significantly in their rotational and magnetic activity properties. GJ 411 is a slowly rotating inactive star, whereas GJ 617A is a moderately rotating flare star. The long-term variability observed in GJ 617A seems to be consistent with a solar-like dynamo mechanism. In contrast, the activity periodicities detected in GJ 411 deviate from the usual semi-empirical relations established for solar-type stars. If these signals are attributed to stellar cycles, this observation would suggest the presence of a different type of dynamo process—likely one in which convective dynamics play a dominant role.

\begin{table*}
\caption{Results of the long-term activity cycle analysis obtained in this work.}
\label{table:cycles}
\centering
\begin{tabular}{p{2.2cm} p{2.2cm} p{1.6cm} p{2cm} p{3cm} p{3cm}}
\hline\hline
Name & $\log R'_{\rm HK}$ & Index & Model & Per1 [days] & Per2 [days] \\
\hline

\multirow{4}{*}{GJ 617A}
& \multirow{4}{*}{-4.85} & $S$       & GLS    & 1831.67 $\pm$ 838.75 \\
&                        & $S$       & LinGLS & 1752.03 $\pm$ 767.40 \\
&                        & H$\alpha$ & GLS    & 1918.93 $\pm$ 920.57 \\
&                        & H$\alpha$ & LinGLS & 1752.07 $\pm$ 767.43 \\
\hline

\multirow{4}{*}{GJ 411}
& \multirow{4}{*}{-5.42} & $S$       & GLS    & 688.02 $\pm$ 118.34  & 386.60 $\pm$ 37.36 \\
&                        & $S$       & LinGLS & 665.47 $\pm$ 110.70  & 2136.48 $\pm$ 1141.13 \\
&                        & H$\alpha$ & GLS    & \multicolumn{2}{c}{not fit} \\
&                        & H$\alpha$ & LinGLS & 1623.70 $\pm$ 659.10 \\
\hline
\end{tabular}

\tablefoot{
Columns list: (1) stellar name; (2) mean $\log R'_{\rm HK}$ value, computed from the mean SOPHIE $S$-index using the calibration of \citet{Boisse2010} for the $V-K$ color and the coefficients of \citet{AstudilloDefru2017}; (3) activity index used in the analysis ($S$ or H$\alpha$); (4) model applied to the data (GLS or LinGLS); and (5)–(6) detected periods in days. Uncertainties correspond to the half width at half maximum of the corresponding peak in the periodogram.
}
\end{table*}

\begin{acknowledgements}

We warmly thank the OHP staff for their support on the
observations. IB  received funding from the French Programme National de Physique Stellaire (PNPS) and the Programme National de Planétologie (PNP) of CNRS (INSU).\\

R.P. and E.J. acknowledge funding from CONICET, under projects PIBAA-CONICET ID-73811 and ID-73669. Funding for the TESS mission is provided by NASA’s Science Mission Directorate. We acknowledge the use of public TESS data from pipelines at the TESS Science Office and at the TESS Science Processing Operations Center. Resources supporting this work were provided by the NASA High-End Computing (HEC) Program through the NASA Advanced Supercomputing (NAS) Division at Ames Research Center for the production of the SPOC data products. Data presented in this paper were obtained from the Mikulski Archive for Space Telescopes (MAST). This work has made use of data from the European Space Agency (ESA) mission Gaia (https://www.cosmos.esa.int/gaia), processed by the Gaia Data Processing and Analysis Consortium (DPAC, https: //www.cosmos.esa.int/web/gaia/dpac/consortium). Funding for the DPAC has been provided by national institutions, in particular the institutions participating in the Gaia Multilateral Agreement.\\

C. G. O. also wishes to express their heartfelt gratitude to Hilda E. Jara for her unconditional support throughout this work, and to C.A.S.L.A., her great passion.
\end{acknowledgements}

\bibliographystyle{aa}
\bibliography{bibliografia}

\begin{thebibliography}{95}
\expandafter\ifx\csname natexlab\endcsname\relax\def\natexlab#1{#1}\fi

\bibitem[{{Adibekyan} {et~al.}(2011){Adibekyan}, {Santos}, {Sousa}, \&
  {Israelian}}]{Adibekyan2011}
{Adibekyan}, V.~Z., {Santos}, N.~C., {Sousa}, S.~G., \& {Israelian}, G. 2011,
  \aap, 535, L11

\bibitem[{{Alonso-Floriano}(2015)}]{Alonso-Floriano2015}
{Alonso-Floriano}, F.~J. 2015, PhD thesis, Complutense University of Madrid,
  Department of Astronomy

\bibitem[{{Anglada-Escud{\'e}} {et~al.}(2016){Anglada-Escud{\'e}}, {Amado},
  {Barnes}, {Berdi{\~n}as}, {Butler}, {Coleman}, {de La Cueva}, {Dreizler},
  {Endl}, {Giesers}, {Jeffers}, {Jenkins}, {Jones}, {Kiraga}, {K{\"u}rster},
  {L{\'o}pez-Gonz{\'a}lez}, {Marvin}, {Morales}, {Morin}, {Nelson}, {Ortiz},
  {Ofir}, {Paardekooper}, {Reiners}, {Rodr{\'\i}guez},
  {Rodr{\'\i}guez-L{\'o}pez}, {Sarmiento}, {Strachan}, {Tsapras}, {Tuomi}, \&
  {Zechmeister}}]{AngladaEscude2016}
{Anglada-Escud{\'e}}, G., {Amado}, P.~J., {Barnes}, J., {et~al.} 2016, \nat,
  536, 437

\bibitem[{{Astudillo-Defru} {et~al.}(2017){Astudillo-Defru}, {Delfosse},
  {Bonfils}, {Forveille}, {Lovis}, \& {Rameau}}]{AstudilloDefru2017}
{Astudillo-Defru}, N., {Delfosse}, X., {Bonfils}, X., {et~al.} 2017, \aap, 600,
  A13

\bibitem[{{Baliunas} {et~al.}(1995){Baliunas}, {Donahue}, {Soon}, {Horne},
  {Frazer}, {Woodard-Eklund}, {Bradford}, {Rao}, {Wilson}, {Zhang}, {Bennett},
  {Briggs}, {Carroll}, {Duncan}, {Figueroa}, {Lanning}, {Misch}, {Mueller},
  {Noyes}, {Poppe}, {Porter}, {Robinson}, {Russell}, {Shelton}, {Soyumer},
  {Vaughan}, \& {Whitney}}]{Baliunas1995}
{Baliunas}, S.~L., {Donahue}, R.~A., {Soon}, W.~H., {et~al.} 1995, \apj, 438,
  269

\bibitem[{{Baycroft} {et~al.}(2025){Baycroft}, {Santerne}, {Triaud}, {Heidari},
  {Sebastian}, {Davis}, {Correia}, {Sairam}, {Freckelton}, {Adamson}, {Boisse},
  {Coleman}, {Dransfield}, {Faria}, {Grouffal}, {Hara}, {H{\'e}brard},
  {Kunovac}, {Martin}, {Maxted}, {Nelson}, {Scott}, {Scutt}, \&
  {Standing}}]{Baycroft2025}
{Baycroft}, T.~A., {Santerne}, A., {Triaud}, A. H.~M.~J., {et~al.} 2025,
  \mnras, 541, 2801

\bibitem[{{Bensby} {et~al.}(2005){Bensby}, {Feltzing}, {Lundstr{\"o}m}, \&
  {Ilyin}}]{Bensby2005}
{Bensby}, T., {Feltzing}, S., {Lundstr{\"o}m}, I., \& {Ilyin}, I. 2005, \aap,
  433, 185

\bibitem[{{Bochanski} {et~al.}(2010){Bochanski}, {Hawley}, {Covey}, {West},
  {Reid}, {Golimowski}, \& {Ivezi{\'c}}}]{Bochanski2010}
{Bochanski}, J.~J., {Hawley}, S.~L., {Covey}, K.~R., {et~al.} 2010, \aj, 139,
  2679

\bibitem[{{Boisse} {et~al.}(2011){Boisse}, {Bouchy}, {H{\'e}brard}, {Bonfils},
  {Santos}, \& {Vauclair}}]{Boisse2011}
{Boisse}, I., {Bouchy}, F., {H{\'e}brard}, G., {et~al.} 2011, \aap, 528, A4

\bibitem[{{Boisse} {et~al.}(2010){Boisse}, {Eggenberger}, {Santos}, {Lovis},
  {Bouchy}, {H{\'e}brard}, {Arnold}, {Bonfils}, {Delfosse}, {Desort},
  {D{\'\i}az}, {Ehrenreich}, {Forveille}, {Gallenne}, {Lagrange}, {Moutou},
  {Udry}, {Pepe}, {Perrier}, {Perruchot}, {Pont}, {Queloz}, {Santerne},
  {S{\'e}gransan}, \& {Vidal-Madjar}}]{Boisse2010}
{Boisse}, I., {Eggenberger}, A., {Santos}, N.~C., {et~al.} 2010, \aap, 523, A88

\bibitem[{{Boisse} {et~al.}(2012){Boisse}, {Pepe}, {Perrier}, {Queloz},
  {Bonfils}, {Bouchy}, {Santos}, {Arnold}, {Beuzit}, {D{\'\i}az}, {Delfosse},
  {Eggenberger}, {Ehrenreich}, {Forveille}, {H{\'e}brard}, {Lagrange}, {Lovis},
  {Mayor}, {Moutou}, {Naef}, {Santerne}, {S{\'e}gransan}, {Sivan}, \&
  {Udry}}]{boisse2012}
{Boisse}, I., {Pepe}, F., {Perrier}, C., {et~al.} 2012, \aap, 545, A55

\bibitem[{{Bonfils} {et~al.}(2013){Bonfils}, {Delfosse}, {Udry}, {Forveille},
  {Mayor}, {Perrier}, {Bouchy}, {Gillon}, {Lovis}, {Pepe}, {Queloz}, {Santos},
  {S{\'e}gransan}, \& {Bertaux}}]{Bonfils2013}
{Bonfils}, X., {Delfosse}, X., {Udry}, S., {et~al.} 2013, \aap, 549, A109

\bibitem[{{Bouchy} {et~al.}(2013){Bouchy}, {D{\'\i}az}, {H{\'e}brard},
  {Arnold}, {Boisse}, {Delfosse}, {Perruchot}, \& {Santerne}}]{Bouchy13}
{Bouchy}, F., {D{\'\i}az}, R.~F., {H{\'e}brard}, G., {et~al.} 2013, \aap, 549,
  A49

\bibitem[{{Bouchy} {et~al.}(2011){Bouchy}, {H{\'e}brard}, {Delfosse}, {Udry},
  {Lagrange}, {Arnold}, {Boisse}, {Bonfils}, {Debondt}, {Diaz}, {Eggenberger},
  {Ehrenreich}, {Forveille}, {Lovis}, {Moutou}, {Pepe}, {Perrier}, {Queloz},
  {Santerne}, {Santos}, \& {S{\'e}gransan}}]{Bouchy2011}
{Bouchy}, F., {H{\'e}brard}, G., {Delfosse}, X., {et~al.} 2011, in EPSC-DPS
  Joint Meeting 2011, Vol. 2011, 240

\bibitem[{{Bouchy} {et~al.}(2009){Bouchy}, {H{\'e}brard}, {Udry}, {Delfosse},
  {Boisse}, {Desort}, {Bonfils}, {Eggenberger}, {Ehrenreich}, {Forveille},
  {Lagrange}, {Le Coroller}, {Lovis}, {Moutou}, {Pepe}, {Perrier}, {Pont},
  {Queloz}, {Santos}, {S{\'e}gransan}, \& {Vidal-Madjar}}]{Bouchy2009}
{Bouchy}, F., {H{\'e}brard}, G., {Udry}, S., {et~al.} 2009, \aap, 505, 853

\bibitem[{{Buccino} {et~al.}(2011){Buccino}, {D{\'\i}az}, {Luoni}, {Abrevaya},
  \& {Mauas}}]{Buccino2011}
{Buccino}, A.~P., {D{\'\i}az}, R.~F., {Luoni}, M.~L., {Abrevaya}, X.~C., \&
  {Mauas}, P. J.~D. 2011, \aj, 141, 34

\bibitem[{{Buccino} {et~al.}(2020){Buccino}, {Sraibman}, {Olivar}, \&
  {Minotti}}]{Buccino2020}
{Buccino}, A.~P., {Sraibman}, L., {Olivar}, P.~M., \& {Minotti}, F.~O. 2020,
  \mnras, 497, 3968

\bibitem[{{Butler} {et~al.}(2017{\natexlab{a}}){Butler}, {Vogt}, {Laughlin},
  {Burt}, {Rivera}, {Tuomi}, {Teske}, {Arriagada}, {Diaz}, {Holden}, \&
  {Keiser}}]{Butler2017}
{Butler}, R.~P., {Vogt}, S.~S., {Laughlin}, G., {et~al.} 2017{\natexlab{a}},
  \aj, 153, 208

\bibitem[{{Butler} {et~al.}(2017{\natexlab{b}}){Butler}, {Vogt}, {Laughlin},
  {Burt}, {Rivera}, {Tuomi}, {Teske}, {Arriagada}, {Diaz}, {Holden}, \&
  {Keiser}}]{Butler17}
{Butler}, R.~P., {Vogt}, S.~S., {Laughlin}, G., {et~al.} 2017{\natexlab{b}},
  \aj, 153, 208

\bibitem[{{Cifuentes} {et~al.}(2020){Cifuentes}, {Caballero},
  {Cortes-Contreras}, {Montes}, {Abellan}, {Dorda}, {Holgado}, {Zapatero
  Osorio}, {Morales}, {Amado}, {Passegger}, {Quirrenbach}, {Reiners}, {Ribas},
  {Sanz-Forcada}, {Schweitzer}, {Seifert}, \& {Solano}}]{Cifuentes2020}
{Cifuentes}, C., {Caballero}, J.~A., {Cortes-Contreras}, M., {et~al.} 2020,
  {VizieR Online Data Catalog: CARMENES input catalogue of M dwarfs. V.
  (Cifuentes+, 2020)}, VizieR On-line Data Catalog: J/A+A/642/A115. Originally
  published in: 2020A\&A...642A.115C

\bibitem[{{Cincunegui} {et~al.}(2007){Cincunegui}, {D{\'\i}az}, \&
  {Mauas}}]{Cincunegui2007}
{Cincunegui}, C., {D{\'\i}az}, R.~F., \& {Mauas}, P.~J.~D. 2007, \aap, 469, 309

\bibitem[{{Cortes-Zuleta} {et~al.}(2024){Cortes-Zuleta}, {Boisse},
  {Ould-Elhkim}, {Wilson}, {Larue}, {Carmona}, {Delfosse}, {Donati},
  {Forveille}, {Moutou}, {Collier}, {Artigau}, {Acuna}, {Altinier},
  {Astudillo-Defru}, {Baruteau}, {Bonfils}, {Cabrit}, {Cadieux}, {Cook},
  {Decocq}, {Diaz}, {Fouque}, {Gomes da Silva}, {Grankin}, {Grouffal}, {Hara},
  {Hebrard}, {Heidari}, {Martins}, {Martioli}, {Maurice}, {Scigliuto}, {Serrano
  Bell}, {Sulis}, {Petit}, \& {Vivien}}]{CortesZuleta2024}
{Cortes-Zuleta}, P., {Boisse}, I., {Ould-Elhkim}, M., {et~al.} 2024, {VizieR
  Online Data Catalog: GL725A RVs and activity indicators (Cortes-Zuleta+,
  2025)}, VizieR On-line Data Catalog: J/A+A/693/A164. Originally published in:
  2025A\&A...693A.164C

\bibitem[{{Davenport}(2016)}]{davenport2016}
{Davenport}, J. R.~A. 2016, \apj, 829, 23

\bibitem[{{Demangeon} {et~al.}(2021){Demangeon}, {Dalal}, {H{\'e}brard},
  {Nsamba}, {Kiefer}, {Camacho}, {Sahlmann}, {Arnold}, {Astudillo-Defru},
  {Bonfils}, {Boisse}, {Bouchy}, {Bourrier}, {Campante}, {Delfosse}, {Deleuil},
  {D{\'\i}az}, {Faria}, {Forveille}, {Hara}, {Heidari}, {Hobson}, {Lopez},
  {Moutou}, {Rey}, {Santerne}, {Sousa}, {Santos}, {Str{\o}m}, {Tsantaki}, \&
  {Udry}}]{demangeon2021}
{Demangeon}, O.~D.~S., {Dalal}, S., {H{\'e}brard}, G., {et~al.} 2021, \aap,
  653, A78

\bibitem[{{D{\'\i}az} {et~al.}(2019){D{\'\i}az}, {Delfosse}, {Hobson},
  {Boisse}, {Astudillo-Defru}, {Bonfils}, {Henry}, {Arnold}, {Bouchy},
  {Bourrier}, {Brugger}, {Dalal}, {Deleuil}, {Demangeon}, {Dolon}, {Dumusque},
  {Forveille}, {Hara}, {H{\'e}brard}, {Kiefer}, {Lopez}, {Mignon}, {Moreau},
  {Mousis}, {Moutou}, {Pepe}, {Perruchot}, {Richaud}, {Santerne}, {Santos},
  {Sottile}, {Stalport}, {S{\'e}gransan}, {Udry}, {Unger}, \&
  {Wilson}}]{Diaz19}
{D{\'\i}az}, R.~F., {Delfosse}, X., {Hobson}, M.~J., {et~al.} 2019, \aap, 625,
  A17

\bibitem[{{D{\'\i}az} {et~al.}(2007){D{\'\i}az}, {Ram{\'\i}rez},
  {Fern{\'a}ndez}, {Gallardo}, {Gieren}, {Ivanov}, {Mauas}, {Minniti},
  {Pietrzynski}, {P{\'e}rez}, {Ru{\'\i}z}, {Udalski}, \& {Zoccali}}]{Diaz2007}
{D{\'\i}az}, R.~F., {Ram{\'\i}rez}, S., {Fern{\'a}ndez}, J.~M., {et~al.} 2007,
  \apj, 660, 850

\bibitem[{{D{\'\i}az} {et~al.}(2016){D{\'\i}az}, {Rey}, {Demangeon},
  {H{\'e}brard}, {Boisse}, {Arnold}, {Astudillo-Defru}, {Beuzit}, {Bonfils},
  {Borgniet}, {Bouchy}, {Bourrier}, {Courcol}, {Deleuil}, {Delfosse},
  {Ehrenreich}, {Forveille}, {Lagrange}, {Mayor}, {Moutou}, {Pepe}, {Queloz},
  {Santerne}, {Santos}, {Sahlmann}, {S{\'e}gransan}, {Udry}, \&
  {Wilson}}]{diaz2016}
{D{\'\i}az}, R.~F., {Rey}, J., {Demangeon}, O., {et~al.} 2016, \aap, 591, A146

\bibitem[{{Donati} {et~al.}(2023){Donati}, {Lehmann}, {Cristofari},
  {Fouqu{\'e}}, {Moutou}, {Charpentier}, {Ould-Elhkim}, {Carmona}, {Delfosse},
  {Artigau}, {Alencar}, {Cadieux}, {Arnold}, {Petit}, {Morin}, {Forveille},
  {Cloutier}, {Doyon}, {H{\'e}brard}, \& {SLS Collaboration}}]{Donati2023}
{Donati}, J.~F., {Lehmann}, L.~T., {Cristofari}, P.~I., {et~al.} 2023, \mnras,
  525, 2015

\bibitem[{{Dumusque} {et~al.}(2011{\natexlab{a}}){Dumusque}, {Lovis},
  {S{\'e}gransan}, {Mayor}, {Udry}, {Benz}, {Bouchy}, {Lo Curto}, {Mordasini},
  {Pepe}, {Queloz}, {Santos}, \& {Naef}}]{Dumusque2011}
{Dumusque}, X., {Lovis}, C., {S{\'e}gransan}, D., {et~al.} 2011{\natexlab{a}},
  \aap, 535, A55

\bibitem[{{Dumusque} {et~al.}(2011{\natexlab{b}}){Dumusque}, {Santos}, {Udry},
  {Lovis}, \& {Bonfils}}]{Dumusque2011model}
{Dumusque}, X., {Santos}, N.~C., {Udry}, S., {Lovis}, C., \& {Bonfils}, X.
  2011{\natexlab{b}}, \aap, 527, A82

\bibitem[{{Flores} {et~al.}(2018){Flores}, {Gonz{\'a}lez}, {Jaque Arancibia},
  {Saffe}, {Buccino}, {L{\'o}pez}, {Iba{\~n}ez Bustos}, \&
  {Miquelarena}}]{Flores2018}
{Flores}, M., {Gonz{\'a}lez}, J.~F., {Jaque Arancibia}, M., {et~al.} 2018,
  \aap, 620, A34

\bibitem[{{Flores Trivigno} {et~al.}(2024){Flores Trivigno}, {Buccino},
  {Gonz{\'a}lez}, {Colombo}, {Gonz{\'a}lez}, {Jaque Arancibia}, {Iba{\~n}ez
  Bustos}, {Saffe}, {Miquelarena}, {Alacoria}, \& {Collado}}]{Flores2024}
{Flores Trivigno}, M., {Buccino}, A.~P., {Gonz{\'a}lez}, E., {et~al.} 2024,
  \aap, 691, L6

\bibitem[{{Fouqu{\'e}} {et~al.}(2023){Fouqu{\'e}}, {Martioli}, {Donati},
  {Lehmann}, {Zaire}, {Bellotti}, {Gaidos}, {Morin}, {Moutou}, {Petit},
  {Alencar}, {Arnold}, {Artigau}, {Cang}, {Carmona}, {Cook},
  {Cort{\'e}s-Zuleta}, {Cristofari}, {Delfosse}, {Doyon}, {H{\'e}brard},
  {Malo}, {Reyl{\'e}}, \& {Usher}}]{Fouque2023}
{Fouqu{\'e}}, P., {Martioli}, E., {Donati}, J.~F., {et~al.} 2023, \aap, 672,
  A52

\bibitem[{{Fuhrmeister} {et~al.}(2020){Fuhrmeister}, {Czesla}, {Hildebrandt},
  {Nagel}, {Schmitt}, {Jeffers}, {Caballero}, {Hintz}, {Johnson},
  {Sch{\"o}fer}, {Zechmeister}, {Reiners}, {Ribas}, {Amado}, {Quirrenbach},
  {Nortmann}, {Bauer}, {B{\'e}jar}, {Cort{\'e}s-Contreras}, {Dreizler},
  {Galad{\'\i}-Enr{\'\i}quez}, {Hatzes}, {Kaminski}, {K{\"u}rster}, {Lafarga},
  \& {Montes}}]{Fuhrmeister20}
{Fuhrmeister}, B., {Czesla}, S., {Hildebrandt}, L., {et~al.} 2020, \aap, 640,
  A52

\bibitem[{{Gaia Collaboration} {et~al.}(2018){Gaia Collaboration}, {Katz},
  {Antoja}, {Romero-G{\'o}mez}, {Drimmel}, {Reyl{\'e}}, {Seabroke}, {Soubiran},
  {Babusiaux}, {Di Matteo}, {Figueras}, {Poggio}, {Robin}, {Evans}, {Brown},
  {Vallenari}, {Prusti}, {de Bruijne}, {Bailer-Jones}, {Biermann}, {Eyer},
  {Jansen}, {Jordi}, {Klioner}, {Lammers}, {Lindegren}, {Luri}, {Mignard},
  {Panem}, {Pourbaix}, {Randich}, {Sartoretti}, {Siddiqui}, {van Leeuwen},
  {Walton}, {Arenou}, {Bastian}, {Cropper}, {Lattanzi}, {Bakker}, {Cacciari},
  {Casta n}, {Chaoul}, {Cheek}, {De Angeli}, {Fabricius}, {Guerra}, {Holl},
  {Masana}, {Messineo}, {Mowlavi}, {Nienartowicz}, {Panuzzo}, {Portell},
  {Riello}, {Tanga}, {Th{\'e}venin}, {Gracia-Abril}, {Comoretto},
  {Garcia-Reinaldos}, {Teyssier}, {Altmann}, {Andrae}, {Audard},
  {Bellas-Velidis}, {Benson}, {Berthier}, {Blomme}, {Burgess}, {Busso},
  {Carry}, {Cellino}, {Clementini}, {Clotet}, {Creevey}, {Davidson}, {De
  Ridder}, {Delchambre}, {Dell'Oro}, {Ducourant},
  {Fern{\'a}ndez-Hern{\'a}ndez}, {Fouesneau}, {Fr{\'e}mat}, {Galluccio},
  {Garc{\'\i}a-Torres}, {Gonz{\'a}lez-N{\'u}{\~n}ez}, {Gonz{\'a}lez-Vidal},
  {Gosset}, {Guy}, {Halbwachs}, {Hambly}, {Harrison}, {Hern{\'a}ndez},
  {Hestroffer}, {Hodgkin}, {Hutton}, {Jasniewicz}, {Jean-Antoine-Piccolo},
  {Jordan}, {Korn}, {Krone-Martins}, {Lanzafame}, {Lebzelter}, {L{\"o}ffler},
  {Manteiga}, {Marrese}, {Mart{\'\i}n-Fleitas}, {Moitinho}, {Mora}, {Muinonen},
  {Osinde}, {Pancino}, {Pauwels}, {Petit}, {Recio-Blanco}, {Richards},
  {Rimoldini}, {Sarro}, {Siopis}, {Smith}, {Sozzetti}, {S{\"u}veges}, {Torra},
  {van Reeven}, {Abbas}, {Abreu Aramburu}, {Accart}, {Aerts}, {Altavilla},
  {{\'A}lvarez}, {Alvarez}, {Alves}, {Anderson}, {Andrei}, {Anglada Varela},
  {Antiche}, {Arcay}, {Astraatmadja}, {Bach}, {Baker},
  {Balaguer-N{\'u}{\~n}ez}, {Balm}, {Barache}, {Barata}, {Barbato}, {Barblan},
  {Barklem}, {Barrado}, {Barros}, {Barstow}, {Bartholom{\'e} Mu{\~n}oz},
  {Bassilana}, {Becciani}, {Bellazzini}, {Berihuete}, {Bertone}, {Bianchi},
  {Bienaym{\'e}}, {Blanco-Cuaresma}, {Boch}, {Boeche}, {Bombrun}, {Borrachero},
  {Bossini}, {Bouquillon}, {Bourda}, {Bragaglia}, {Bramante}, {Breddels},
  {Bressan}, {Brouillet}, {Br{\"u}semeister}, {Brugaletta}, {Bucciarelli},
  {Burlacu}, {Busonero}, {Butkevich}, {Buzzi}, {Caffau}, {Cancelliere},
  {Cannizzaro}, {Cantat-Gaudin}, {Carballo}, {Carlucci}, {Carrasco},
  {Casamiquela}, {Castellani}, {Castro-Ginard}, {Charlot}, {Chemin},
  {Chiavassa}, {Cocozza}, {Costigan}, {Cowell}, {Crifo}, {Crosta}, {Crowley},
  {Cuypers}, {Dafonte}, \& {Damerdji}}]{GaiaCollaboration2018}
{Gaia Collaboration}, {Katz}, D., {Antoja}, T., {et~al.} 2018, \aap, 616, A11

\bibitem[{{Gaia Collaboration} {et~al.}(2023){Gaia Collaboration}, {Vallenari},
  {Brown}, {Prusti}, {de Bruijne}, {Arenou}, {Babusiaux}, {Biermann},
  {Creevey}, {Ducourant}, {Evans}, {Eyer}, {Guerra}, {Hutton}, {Jordi},
  {Klioner}, {Lammers}, {Lindegren}, {Luri}, {Mignard}, {Panem}, {Pourbaix},
  {Randich}, {Sartoretti}, {Soubiran}, {Tanga}, {Walton}, {Bailer-Jones},
  {Bastian}, {Drimmel}, {Jansen}, {Katz}, {Lattanzi}, {van Leeuwen}, {Bakker},
  {Cacciari}, {Casta{\~n}eda}, {De Angeli}, {Fabricius}, {Fouesneau},
  {Fr{\'e}mat}, {Galluccio}, {Guerrier}, {Heiter}, {Masana}, {Messineo},
  {Mowlavi}, {Nicolas}, {Nienartowicz}, {Pailler}, {Panuzzo}, {Riclet}, {Roux},
  {Seabroke}, {Sordo}, {Th{\'e}venin}, {Gracia-Abril}, {Portell}, {Teyssier},
  {Altmann}, {Andrae}, {Audard}, {Bellas-Velidis}, {Benson}, {Berthier},
  {Blomme}, {Burgess}, {Busonero}, {Busso}, {C{\'a}novas}, {Carry}, {Cellino},
  {Cheek}, {Clementini}, {Damerdji}, {Davidson}, {de Teodoro}, {Nu{\~n}ez
  Campos}, {Delchambre}, {Dell'Oro}, {Esquej}, {Fern{\'a}ndez-Hern{\'a}ndez},
  {Fraile}, {Garabato}, {Garc{\'\i}a-Lario}, {Gosset}, {Haigron}, {Halbwachs},
  {Hambly}, {Harrison}, {Hern{\'a}ndez}, {Hestroffer}, {Hodgkin}, {Holl},
  {Jan{\ss}en}, {Jevardat de Fombelle}, {Jordan}, {Krone-Martins}, {Lanzafame},
  {L{\"o}ffler}, {Marchal}, {Marrese}, {Moitinho}, {Muinonen}, {Osborne},
  {Pancino}, {Pauwels}, {Recio-Blanco}, {Reyl{\'e}}, {Riello}, {Rimoldini},
  {Roegiers}, {Rybizki}, {Sarro}, {Siopis}, {Smith}, {Sozzetti}, {Utrilla},
  {van Leeuwen}, {Abbas}, {{\'A}brah{\'a}m}, {Abreu Aramburu}, {Aerts},
  {Aguado}, {Ajaj}, {Aldea-Montero}, {Altavilla}, {{\'A}lvarez}, {Alves},
  {Anders}, {Anderson}, {Anglada Varela}, {Antoja}, {Baines}, {Baker},
  {Balaguer-N{\'u}{\~n}ez}, {Balbinot}, {Balog}, {Barache}, {Barbato},
  {Barros}, {Barstow}, {Bartolom{\'e}}, {Bassilana}, {Bauchet}, {Becciani},
  {Bellazzini}, {Berihuete}, {Bernet}, {Bertone}, {Bianchi}, {Binnenfeld},
  {Blanco-Cuaresma}, {Blazere}, {Boch}, {Bombrun}, {Bossini}, {Bouquillon},
  {Bragaglia}, {Bramante}, {Breedt}, {Bressan}, {Brouillet}, {Brugaletta},
  {Bucciarelli}, {Burlacu}, {Butkevich}, {Buzzi}, {Caffau}, {Cancelliere},
  {Cantat-Gaudin}, {Carballo}, {Carlucci}, {Carnerero}, {Carrasco},
  {Casamiquela}, {Castellani}, {Castro-Ginard}, {Chaoul}, {Charlot}, {Chemin},
  {Chiaramida}, {Chiavassa}, {Chornay}, {Comoretto}, {Contursi}, {Cooper},
  {Cornez}, {Cowell}, {Crifo}, {Cropper}, {Crosta}, {Crowley}, {Dafonte},
  {Dapergolas}, {David}, {David}, {de Laverny}, {De Luise}, \& {De
  March}}]{GaiaCollaboration2023}
{Gaia Collaboration}, {Vallenari}, A., {Brown}, A.~G.~A., {et~al.} 2023, \aap,
  674, A1

\bibitem[{{Gizis} {et~al.}(2002){Gizis}, {Reid}, \& {Hawley}}]{Gizis2002}
{Gizis}, J.~E., {Reid}, I.~N., \& {Hawley}, S.~L. 2002, \aj, 123, 3356

\bibitem[{{Gomes da Silva} {et~al.}(2022){Gomes da Silva}, {Bensabat},
  {Monteiro}, \& {Santos}}]{GomesdaSilva22}
{Gomes da Silva}, J., {Bensabat}, A., {Monteiro}, T., \& {Santos}, N.~C. 2022,
  \aap, 668, A174

\bibitem[{{Gomes da Silva} {et~al.}(2014){Gomes da Silva}, {Santos}, {Boisse},
  {Dumusque}, \& {Lovis}}]{GomesDaSilva2014}
{Gomes da Silva}, J., {Santos}, N.~C., {Boisse}, I., {Dumusque}, X., \&
  {Lovis}, C. 2014, \aap, 566, A66

\bibitem[{{Gomes da Silva} {et~al.}(2011){Gomes da Silva}, {Santos}, {Bonfils},
  {Delfosse}, {Forveille}, \& {Udry}}]{GomesdaSilva2011}
{Gomes da Silva}, J., {Santos}, N.~C., {Bonfils}, X., {et~al.} 2011, \aap, 534,
  A30

\bibitem[{{Gomes da Silva} {et~al.}(2012){Gomes da Silva}, {Santos}, {Bonfils},
  {Delfosse}, {Forveille}, {Udry}, {Dumusque}, \& {Lovis}}]{GomesdaSilva2012}
{Gomes da Silva}, J., {Santos}, N.~C., {Bonfils}, X., {et~al.} 2012, \aap, 541,
  A9

\bibitem[{{Hall} {et~al.}(2009){Hall}, {Henry}, {Lockwood}, {Skiff}, \&
  {Saar}}]{Hall2009}
{Hall}, J.~C., {Henry}, G.~W., {Lockwood}, G.~W., {Skiff}, B.~A., \& {Saar},
  S.~H. 2009, \aj, 138, 312

\bibitem[{{Hara} {et~al.}(2020){Hara}, {Bouchy}, {Stalport}, {Boisse},
  {Rodrigues}, {Delisle}, {Santerne}, {Henry}, {Arnold}, {Astudillo-Defru},
  {Borgniet}, {Bonfils}, {Bourrier}, {Brugger}, {Courcol}, {Dalal}, {Deleuil},
  {Delfosse}, {Demangeon}, {D{\'\i}az}, {Dumusque}, {Forveille}, {H{\'e}brard},
  {Hobson}, {Kiefer}, {Lopez}, {Mignon}, {Mousis}, {Moutou}, {Pepe}, {Rey},
  {Santos}, {S{\'e}gransan}, {Udry}, \& {Wilson}}]{hara2020}
{Hara}, N.~C., {Bouchy}, F., {Stalport}, M., {et~al.} 2020, \aap, 636, L6

\bibitem[{{H{\'e}brard} {et~al.}(2010){H{\'e}brard}, {Bonfils},
  {S{\'e}gransan}, {Moutou}, {Delfosse}, {Bouchy}, {Boisse}, {Arnold},
  {Desort}, {D{\'\i}az}, {Eggenberger}, {Ehrenreich}, {Forveille}, {Lagrange},
  {Lovis}, {Pepe}, {Perrier}, {Pont}, {Queloz}, {Santos}, {Udry}, \&
  {Vidal-Madjar}}]{hebrard2010}
{H{\'e}brard}, G., {Bonfils}, X., {S{\'e}gransan}, D., {et~al.} 2010, \aap,
  513, A69

\bibitem[{{Heidari} {et~al.}(2024){Heidari}, {Boisse}, {Hara}, {Wilson},
  {Kiefer}, {H{\'e}brard}, {Philipot}, {Hoyer}, {Stassun}, {Henry}, {Santos},
  {Acu{\~n}a}, {Almasian}, {Arnold}, {Astudillo-Defru}, {Attia}, {Bonfils},
  {Bouchy}, {Bourrier}, {Collet}, {Cort{\'e}s-Zuleta}, {Carmona}, {Delfosse},
  {Dalal}, {Deleuil}, {Demangeon}, {D{\'\i}az}, {Dumusque}, {Ehrenreich},
  {Forveille}, {Hobson}, {Jenkins}, {Jenkins}, {Lagrange}, {Latham}, {Larue},
  {Liu}, {Moutou}, {Mignon}, {Osborn}, {Pepe}, {Rapetti}, {Rodrigues},
  {Santerne}, {Segransan}, {Shporer}, {Sulis}, {Torres}, {Udry}, {Vakili},
  {Vanderburg}, {Venot}, {Vivien}, \& {Vines}}]{heidari2024}
{Heidari}, N., {Boisse}, I., {Hara}, N.~C., {et~al.} 2024, \aap, 681, A55

\bibitem[{{Hobson} {et~al.}(2018){Hobson}, {D{\'\i}az}, {Delfosse},
  {Astudillo-Defru}, {Boisse}, {Bouchy}, {Bonfils}, {Forveille}, {Hara},
  {Arnold}, {Borgniet}, {Bourrier}, {Brugger}, {Cabrera}, {Courcol}, {Dalal},
  {Deleuil}, {Demangeon}, {Dumusque}, {Ehrenreich}, {H{\'e}brard}, {Kiefer},
  {Lopez}, {Mignon}, {Montagnier}, {Mousis}, {Moutou}, {Pepe}, {Rey},
  {Santerne}, {Santos}, {Stalport}, {S{\'e}gransan}, {Udry}, \&
  {Wilson}}]{Hobson2018}
{Hobson}, M.~J., {D{\'\i}az}, R.~F., {Delfosse}, X., {et~al.} 2018, \aap, 618,
  A103

\bibitem[{{Hurt} {et~al.}(2022){Hurt}, {Fulton}, {Isaacson}, {Rosenthal},
  {Howard}, {Weiss}, \& {Petigura}}]{Hurt2022}
{Hurt}, S.~A., {Fulton}, B., {Isaacson}, H., {et~al.} 2022, \aj, 163, 218

\bibitem[{{Iba{\~n}ez Bustos} {et~al.}(2023){Iba{\~n}ez Bustos}, {Buccino},
  {Flores}, {Martinez}, \& {Mauas}}]{IbanezBustos23}
{Iba{\~n}ez Bustos}, R.~V., {Buccino}, A.~P., {Flores}, M., {Martinez}, C.~F.,
  \& {Mauas}, P.~J.~D. 2023, \aap, 672, A37

\bibitem[{{Iba{\~n}ez Bustos} {et~al.}(2019){Iba{\~n}ez Bustos}, {Buccino},
  {Flores}, \& {Mauas}}]{IbanezBustos2019b}
{Iba{\~n}ez Bustos}, R.~V., {Buccino}, A.~P., {Flores}, M., \& {Mauas},
  P.~J.~D. 2019, \aap, 628, L1

\bibitem[{{Iba{\~n}ez Bustos} {et~al.}(2020){Iba{\~n}ez Bustos}, {Buccino},
  {Messina}, {Lanza}, \& {Mauas}}]{IbanezBustos2020}
{Iba{\~n}ez Bustos}, R.~V., {Buccino}, A.~P., {Messina}, S., {Lanza}, A.~F., \&
  {Mauas}, P.~J.~D. 2020, \aap, 644, A2

\bibitem[{{Iba{\~n}ez Bustos} {et~al.}(2025){Iba{\~n}ez Bustos}, {Buccino},
  {Nardetto}, {Mourard}, {Flores}, \& {Mauas}}]{IbanezBustos2025}
{Iba{\~n}ez Bustos}, R.~V., {Buccino}, A.~P., {Nardetto}, N., {et~al.} 2025,
  \aap, 696, A230

\bibitem[{{Ilin} {et~al.}(2021){Ilin}, {Schmidt}, {Poppenh{\"a}ger},
  {Davenport}, {Kristiansen}, \& {Omohundro}}]{ilin2021}
{Ilin}, E., {Schmidt}, S.~J., {Poppenh{\"a}ger}, K., {et~al.} 2021, \aap, 645,
  A42

\bibitem[{{Jeffers} {et~al.}(2018){Jeffers}, {Sch{\"o}fer}, {Lamert},
  {Reiners}, {Montes}, {Caballero}, {Cort{\'e}s-Contreras}, {Marvin},
  {Passegger}, {Zechmeister}, {Quirrenbach}, {Alonso-Floriano}, {Amado},
  {Bauer}, {Casal}, {Diez Alonso}, {Herrero}, {Morales}, {Mundt}, {Ribas}, \&
  {Sarmiento}}]{Jeffers2018}
{Jeffers}, S.~V., {Sch{\"o}fer}, P., {Lamert}, A., {et~al.} 2018, \aap, 614,
  A76

\bibitem[{{Jenkins} {et~al.}(2016){Jenkins}, {Twicken}, {McCauliff},
  {Campbell}, {Sanderfer}, {Lung}, {Mansouri-Samani}, {Girouard}, {Tenenbaum},
  {Klaus}, {Smith}, {Caldwell}, {Chacon}, {Henze}, {Heiges}, {Latham},
  {Morgan}, {Swade}, {Rinehart}, \& {Vanderspek}}]{jenkins2016}
{Jenkins}, J.~M., {Twicken}, J.~D., {McCauliff}, S., {et~al.} 2016, in Society
  of Photo-Optical Instrumentation Engineers (SPIE) Conference Series, Vol.
  9913, Software and Cyberinfrastructure for Astronomy IV, ed. G.~{Chiozzi} \&
  J.~C. {Guzman}, 99133E

\bibitem[{{Jofr{\'e}} {et~al.}(2015){Jofr{\'e}}, {Petrucci}, {Saffe}, {Saker},
  {Artur de la Villarmois}, {Chavero}, {G{\'o}mez}, \& {Mauas}}]{Jofre2015}
{Jofr{\'e}}, E., {Petrucci}, R., {Saffe}, C., {et~al.} 2015, \aap, 574, A50

\bibitem[{{Kar} {et~al.}(2024){Kar}, {Henry}, {Couperus}, {Vrijmoet}, \&
  {Jao}}]{RECONS2023}
{Kar}, A., {Henry}, T.~J., {Couperus}, A.~A., {Vrijmoet}, E.~H., \& {Jao},
  W.~C. 2024, {VizieR Online Data Catalog: The solar neighborhood LI. RECONS \&
  TESS obs. (Kar+, 2024)}, VizieR On-line Data Catalog: J/AJ/167/196.
  Originally published in: 2024AJ....167..196K

\bibitem[{{Kopparapu} {et~al.}(2013){Kopparapu}, {Ramirez}, {Kasting}, {Eymet},
  {Robinson}, {Mahadevan}, {Terrien}, {Domagal-Goldman}, {Meadows}, \&
  {Deshpande}}]{Kopparapu2013}
{Kopparapu}, R.~K., {Ramirez}, R., {Kasting}, J.~F., {et~al.} 2013, \apj, 765,
  131

\bibitem[{{Lafarga} {et~al.}(2021){Lafarga}, {Ribas}, {Reiners}, {Quirrenbach},
  {Amado}, {Caballero}, {Azzaro}, {Bejar}, {Cortes-Contreras}, {Dreizler},
  {Hatzes}, {Henning}, {Jeffers}, {Kaminski}, {Kuerster}, {Montes}, {Morales},
  {Oshagh}, {Rodriguez-Lopez}, {Schoefer}, {Schweitzer}, \&
  {Zechmeister}}]{Lafarga2021}
{Lafarga}, M., {Ribas}, I., {Reiners}, A., {et~al.} 2021, {VizieR Online Data
  Catalog: Activity indicators across the M dwarf domain (Lafarga+, 2021)},
  VizieR On-line Data Catalog: J/A+A/652/A28. Originally published in:
  2021A\&A...652A..28L

\bibitem[{{Leenaarts} {et~al.}(2012){Leenaarts}, {Carlsson}, \& {Rouppe van der
  Voort}}]{Leenaarts2012}
{Leenaarts}, J., {Carlsson}, M., \& {Rouppe van der Voort}, L. 2012, \apj, 749,
  136

\bibitem[{{Lehmann} {et~al.}(2024){Lehmann}, {Donati}, {Fouqu{\'e}}, {Moutou},
  {Bellotti}, {Delfosse}, {Petit}, {Carmona}, {Morin}, {Vidotto}, \& {the SLS
  consortium}}]{Lehmann2024}
{Lehmann}, L.~T., {Donati}, J.~F., {Fouqu{\'e}}, P., {et~al.} 2024, \mnras,
  527, 4330

\bibitem[{{Lightkurve Collaboration} {et~al.}(2018){Lightkurve Collaboration},
  {Cardoso}, {Hedges}, {Gully-Santiago}, {Saunders}, {Cody}, {Barclay}, {Hall},
  {Sagear}, {Turtelboom}, {Zhang}, {Tzanidakis}, {Mighell}, {Coughlin}, {Bell},
  {Berta-Thompson}, {Williams}, {Dotson}, \& {Barentsen}}]{lightkurve2018}
{Lightkurve Collaboration}, {Cardoso}, J.~V.~d.~M., {Hedges}, C., {et~al.}
  2018, {Lightkurve: Kepler and TESS time series analysis in Python},
  Astrophysics Source Code Library

\bibitem[{{Lovis} {et~al.}(2011){Lovis}, {S{\'e}gransan}, {Mayor}, {Udry},
  {Benz}, {Bertaux}, {Bouchy}, {Correia}, {Laskar}, {Lo Curto}, {Mordasini},
  {Pepe}, {Queloz}, \& {Santos}}]{Lovis2011}
{Lovis}, C., {S{\'e}gransan}, D., {Mayor}, M., {et~al.} 2011, \aap, 528, A112

\bibitem[{{Martin} {et~al.}(2017){Martin}, {Fuhrmeister}, {Mittag}, {Schmidt},
  {Hempelmann}, {Gonz{\'a}lez-P{\'e}rez}, \& {Schmitt}}]{Martin17}
{Martin}, J., {Fuhrmeister}, B., {Mittag}, M., {et~al.} 2017, \aap, 605, A113

\bibitem[{{Martioli} {et~al.}(2024){Martioli}, {Petrucci}, {Jofr{\'e}},
  {H{\'e}brard}, {Ghezzi}, {G{\'o}mez Maqueo Chew}, {D{\'\i}az}, {Perottoni},
  {Garcia}, {Rapetti}, {Lecavelier des Etangs}, {de Almeida}, {Arnold},
  {Artigau}, {Basant}, {Bean}, {Bieryla}, {Boisse}, {Bonfils}, {Brady},
  {Cadieux}, {Carmona}, {Cook}, {Delfosse}, {Donati}, {Doyon}, {Furlan},
  {Howell}, {Jenkins}, {Kasper}, {Kiefer}, {Latham}, {Levine},
  {Lorenzo-Oliveira}, {Luque}, {McLeod}, {Melendez}, {Moutou}, {Netto},
  {Pritchard}, {Rowden}, {Seifahrt}, {Stef{\'a}nsson}, {St{\"u}rmer}, \&
  {Twicken}}]{martioli2024}
{Martioli}, E., {Petrucci}, R.~P., {Jofr{\'e}}, E., {et~al.} 2024, \aap, 690,
  A312

\bibitem[{{Mauas} \& {Falchi}(1994)}]{Mauas1994}
{Mauas}, P. J.~D. \& {Falchi}, A. 1994, \aap, 281, 129

\bibitem[{{Mauas} \& {Falchi}(1996)}]{Mauas1996}
{Mauas}, P.~J.~D. \& {Falchi}, A. 1996, \aap, 310, 245

\bibitem[{{Metcalfe} {et~al.}(2016){Metcalfe}, {Egeland}, \& {van
  Saders}}]{Metcalfe2016}
{Metcalfe}, T.~S., {Egeland}, R., \& {van Saders}, J. 2016, \apjl, 826, L2

\bibitem[{{Meunier} {et~al.}(2010){Meunier}, {Desort}, \&
  {Lagrange}}]{Meunier2010}
{Meunier}, N., {Desort}, M., \& {Lagrange}, A.~M. 2010, \aap, 512, A39

\bibitem[{{Meunier} {et~al.}(2024){Meunier}, {Mignon}, {Kretzschmar}, \&
  {Delfosse}}]{Meunier2024}
{Meunier}, N., {Mignon}, L., {Kretzschmar}, M., \& {Delfosse}, X. 2024, \aap,
  684, A106

\bibitem[{{Mignon} {et~al.}(2023){Mignon}, {Meunier}, {Delfosse}, {Bonfils},
  {Santos}, {Forveille}, {Gaisn{\'e}}, {Astudillo-Defru}, {Lovis}, \&
  {Udry}}]{Mignon2023}
{Mignon}, L., {Meunier}, N., {Delfosse}, X., {et~al.} 2023, \aap, 675, A168

\bibitem[{{Moutou} {et~al.}(2014){Moutou}, {H{\'e}brard}, {Bouchy}, {Arnold},
  {Santos}, {Astudillo-Defru}, {Boisse}, {Bonfils}, {Borgniet}, {Delfosse},
  {D{\'\i}az}, {Ehrenreich}, {Forveille}, {Gregorio}, {Labrevoir}, {Lagrange},
  {Montagnier}, {Montalto}, {Pepe}, {Sahlmann}, {Santerne}, {S{\'e}gransan},
  {Udry}, \& {Vanhuysse}}]{moutou2014}
{Moutou}, C., {H{\'e}brard}, G., {Bouchy}, F., {et~al.} 2014, \aap, 563, A22

\bibitem[{{Newton} {et~al.}(2017){Newton}, {Irwin}, {Charbonneau}, {Berlind},
  {Calkins}, \& {Mink}}]{Newton2017}
{Newton}, E.~R., {Irwin}, J., {Charbonneau}, D., {et~al.} 2017, \apj, 834, 85

\bibitem[{{Newton} {et~al.}(2016){Newton}, {Irwin}, {Charbonneau},
  {Berta-Thompson}, \& {Dittmann}}]{Newton2016}
{Newton}, E.~R., {Irwin}, J., {Charbonneau}, D., {Berta-Thompson}, Z.~K., \&
  {Dittmann}, J.~A. 2016, \apjl, 821, L19

\bibitem[{{Perruchot} {et~al.}(2011){Perruchot}, {Bouchy}, {Chazelas},
  {D{\'\i}az}, {H{\'e}brard}, {Arnaud}, {Arnold}, {Avila}, {Delfosse},
  {Boisse}, {Moreaux}, {Pepe}, {Richaud}, {Santerne}, {Sottile}, \&
  {T{\'e}zier}}]{Perruchot11}
{Perruchot}, S., {Bouchy}, F., {Chazelas}, B., {et~al.} 2011, in Society of
  Photo-Optical Instrumentation Engineers (SPIE) Conference Series, Vol. 8151,
  Techniques and Instrumentation for Detection of Exoplanets V, ed.
  S.~{Shaklan}, 815115

\bibitem[{{Perruchot} {et~al.}(2008){Perruchot}, {Kohler}, {Bouchy}, {Richaud},
  {Richaud}, {Moreaux}, {Merzougui}, {Sottile}, {Hill}, {Knispel}, {Regal},
  {Meunier}, {Ilovaisky}, {Le Coroller}, {Gillet}, {Schmitt}, {Pepe}, {Fleury},
  {Sosnowska}, {Vors}, {M{\'e}gevand}, {Blanc}, {Carol}, {Point}, {Laloge}, \&
  {Brunel}}]{Perruchot2008}
{Perruchot}, S., {Kohler}, D., {Bouchy}, F., {et~al.} 2008, in Society of
  Photo-Optical Instrumentation Engineers (SPIE) Conference Series, Vol. 7014,
  Ground-based and Airborne Instrumentation for Astronomy II, ed. I.~S.
  {McLean} \& M.~M. {Casali}, 70140J

\bibitem[{{Petrucci} {et~al.}(2024){Petrucci}, {G{\'o}mez Maqueo Chew},
  {Jofr{\'e}}, {Segura}, \& {Ferrero}}]{petrucci2024}
{Petrucci}, R.~P., {G{\'o}mez Maqueo Chew}, Y., {Jofr{\'e}}, E., {Segura}, A.,
  \& {Ferrero}, L.~V. 2024, \mnras, 527, 8290

\bibitem[{{Queloz} {et~al.}(2009){Queloz}, {Bouchy}, {Moutou}, {Hatzes},
  {Hebrard}, {Alonso}, {Auvergne}, {Baglin}, {Barbieri}, {Barge}, {Benz},
  {Borde}, {Deeg}, {Deleuil}, {Dvorak}, {Erikson}, {Ferraz Mello}, {Fridlund},
  {Gandolfi}, {Gillon}, {Guenther}, {Guillot}, {Jorda}, {Hartmann}, {Lammer},
  {Leger}, {Llebaria}, {Lovis}, {Magain}, {Mayor}, {Mazeh}, {Ollivier},
  {Patzold}, {Pepe}, {Rauer}, {Rouan}, {Schneider}, {Segransan}, {Udry}, \&
  {Wuchterl}}]{Queloz2009}
{Queloz}, D., {Bouchy}, F., {Moutou}, C., {et~al.} 2009, {VizieR Online Data
  Catalog: CoRoT-7 radial velocities (Queloz+, 2009)}, VizieR On-line Data
  Catalog: J/A+A/506/303. Originally published in: 2009A\&A...506..303Q

\bibitem[{{Rajpaul} {et~al.}(2015){Rajpaul}, {Aigrain}, {Osborne}, {Reece}, \&
  {Roberts}}]{Rajpaul2015}
{Rajpaul}, V., {Aigrain}, S., {Osborne}, M.~A., {Reece}, S., \& {Roberts}, S.
  2015, \mnras, 452, 2269

\bibitem[{{Reddy} {et~al.}(2006){Reddy}, {Lambert}, \& {Allende
  Prieto}}]{Reddy2006}
{Reddy}, B.~E., {Lambert}, D.~L., \& {Allende Prieto}, C. 2006, \mnras, 367,
  1329

\bibitem[{{Reiners} {et~al.}(2018){Reiners}, {Ribas}, {Zechmeister},
  {Caballero}, {Trifonov}, {Dreizler}, {Morales}, {Tal-Or}, {Lafarga},
  {Quirrenbach}, {Amado}, {Kaminski}, {Jeffers}, {Aceituno}, {B{\'e}jar},
  {Gu{\`a}rdia}, {Guenther}, {Hagen}, {Montes}, {Passegger}, {Seifert},
  {Schweitzer}, {Cort{\'e}s-Contreras}, {Abril}, {Alonso-Floriano}, {Ammler-von
  Eiff}, {Antona}, {Anglada-Escud{\'e}}, {Anwand-Heerwart}, {Arroyo-Torres},
  {Azzaro}, {Baroch}, {Barrado}, {Bauer}, {Becerril}, {Ben{\'\i}tez},
  {Berdi{\~n}as}, {Bergond}, {Bl{\"u}mcke}, {Brinkm{\"o}ller}, {del Burgo},
  {Cano}, {C{\'a}rdenas V{\'a}zquez}, {Casal}, {Cifuentes}, {Claret},
  {Colom{\'e}}, {Czesla}, {D{\'\i}ez-Alonso}, {Feiz}, {Fern{\'a}ndez}, {Ferro},
  {Fuhrmeister}, {Galad{\'\i}-Enr{\'\i}quez}, {Garcia-Piquer}, {Garc{\'\i}a
  Vargas}, {Gesa}, {G{\'o}mez Galera}, {Gonz{\'a}lez Hern{\'a}ndez},
  {Gonz{\'a}lez-Peinado}, {Gr{\"o}zinger}, {Grohnert}, {Guijarro}, {de
  Guindos}, {Guti{\'e}rrez-Soto}, {Hatzes}, {Hauschildt}, {Hedrosa},
  {Helmling}, {Henning}, {Hermelo}, {Hern{\'a}ndez Arab{\'\i}}, {Hern{\'a}ndez
  Casta{\~n}o}, {Hern{\'a}ndez Hernando}, {Herrero}, {Huber}, {Huke},
  {Johnson}, {de Juan}, {Kim}, {Klein}, {Kl{\"u}ter}, {Klutsch}, {K{\"u}rster},
  {Labarga}, {Lamert}, {Lamp{\'o}n}, {Lara}, {Laun}, {Lemke}, {Lenzen},
  {Launhardt}, {L{\'o}pez del Fresno}, {L{\'o}pez-Gonz{\'a}lez},
  {L{\'o}pez-Puertas}, {L{\'o}pez Salas}, {L{\'o}pez-Santiago}, {Luque},
  {Mag{\'a}n Madinabeitia}, {Mall}, {Mancini}, {Mandel}, {Marfil}, {Mar{\'\i}n
  Molina}, {Maroto Fern{\'a}ndez}, {Mart{\'\i}n}, {Mart{\'\i}n-Ruiz}, {Marvin},
  {Mathar}, {Mirabet}, {Moreno-Raya}, {Moya}, {Mundt}, {Nagel}, {Naranjo},
  {Nortmann}, {Nowak}, {Ofir}, {Oreiro}, {Pall{\'e}}, {Panduro}, {Pascual},
  {Pavlov}, {Pedraz}, {P{\'e}rez-Calpena}, {P{\'e}rez Medialdea}, {Perger},
  {Perryman}, {Pluto}, {Rabaza}, {Ram{\'o}n}, {Rebolo}, {Redondo}, {Reffert},
  {Reinhart}, {Rhode}, {Rix}, {Rodler}, {Rodr{\'\i}guez},
  {Rodr{\'\i}guez-L{\'o}pez}, {Rodr{\'\i}guez Trinidad}, {Rohloff}, {Rosich},
  {Sadegi}, {S{\'a}nchez-Blanco}, {S{\'a}nchez Carrasco},
  {S{\'a}nchez-L{\'o}pez}, {Sanz-Forcada}, {Sarkis}, {Sarmiento},
  {Sch{\"a}fer}, {Schmitt}, {Schiller}, {Sch{\"o}fer}, {Solano}, {Stahl},
  {Strachan}, {St{\"u}rmer}, {Su{\'a}rez}, {Tabernero}, {Tala}, {Tulloch},
  {Ulbrich}, {Veredas}, {Vico Linares}, {Vilardell}, {Wagner}, {Winkler},
  {Wolthoff}, {Xu}, {Yan}, \& {Zapatero Osorio}}]{Reiners2018}
{Reiners}, A., {Ribas}, I., {Zechmeister}, M., {et~al.} 2018, \aap, 609, L5

\bibitem[{{Reiners} {et~al.}(2014){Reiners}, {Sch{\"u}ssler}, \&
  {Passegger}}]{Reiners2014}
{Reiners}, A., {Sch{\"u}ssler}, M., \& {Passegger}, V.~M. 2014, \apj, 794, 144

\bibitem[{{Ricker} {et~al.}(2015){Ricker}, {Winn}, {Vanderspek}, {Latham},
  {Bakos}, {Bean}, {Berta-Thompson}, {Brown}, {Buchhave}, {Butler}, {Butler},
  {Chaplin}, {Charbonneau}, {Christensen-Dalsgaard}, {Clampin}, {Deming},
  {Doty}, {De Lee}, {Dressing}, {Dunham}, {Endl}, {Fressin}, {Ge}, {Henning},
  {Holman}, {Howard}, {Ida}, {Jenkins}, {Jernigan}, {Johnson}, {Kaltenegger},
  {Kawai}, {Kjeldsen}, {Laughlin}, {Levine}, {Lin}, {Lissauer}, {MacQueen},
  {Marcy}, {McCullough}, {Morton}, {Narita}, {Paegert}, {Palle}, {Pepe},
  {Pepper}, {Quirrenbach}, {Rinehart}, {Sasselov}, {Sato}, {Seager},
  {Sozzetti}, {Stassun}, {Sullivan}, {Szentgyorgyi}, {Torres}, {Udry}, \&
  {Villasenor}}]{ricker2015}
{Ricker}, G.~R., {Winn}, J.~N., {Vanderspek}, R., {et~al.} 2015, Journal of
  Astronomical Telescopes, Instruments, and Systems, 1, 014003

\bibitem[{{Robertson} {et~al.}(2013){Robertson}, {Endl}, {Cochran}, \&
  {Dodson-Robinson}}]{Robertson2013}
{Robertson}, P., {Endl}, M., {Cochran}, W.~D., \& {Dodson-Robinson}, S.~E.
  2013, \apj, 764, 3

\bibitem[{{Robertson} {et~al.}(2014){Robertson}, {Mahadevan}, {Endl}, \&
  {Roy}}]{Robertson2014}
{Robertson}, P., {Mahadevan}, S., {Endl}, M., \& {Roy}, A. 2014, Science, 345,
  440

\bibitem[{{Rodr{\'\i}guez Mart{\'\i}nez} {et~al.}(2020){Rodr{\'\i}guez
  Mart{\'\i}nez}, {Lopez}, {Shappee}, {Schmidt}, {Jayasinghe}, {Kochanek},
  {Auchettl}, \& {Holoien}}]{RodriguezMartinez2020}
{Rodr{\'\i}guez Mart{\'\i}nez}, R., {Lopez}, L.~A., {Shappee}, B.~J., {et~al.}
  2020, \apj, 892, 144

\bibitem[{{Rosenthal} {et~al.}(2021){Rosenthal}, {Fulton}, {Hirsch},
  {Isaacson}, {Howard}, {Dedrick}, {Sherstyuk}, {Blunt}, {Petigura}, {Knutson},
  {Behmard}, {Chontos}, {Crepp}, {Crossfield}, {Dalba}, {Fischer}, {Henry},
  {Kane}, {Kosiarek}, {Marcy}, {Rubenzahl}, {Weiss}, \&
  {Wright}}]{Rosenthal2021}
{Rosenthal}, L.~J., {Fulton}, B.~J., {Hirsch}, L.~A., {et~al.} 2021, \apjs,
  255, 8

\bibitem[{{Saar} \& {Donahue}(1997)}]{Saar1997}
{Saar}, S.~H. \& {Donahue}, R.~A. 1997, \apj, 485, 319

\bibitem[{{Snellen} {et~al.}(2015){Snellen}, {de Kok}, {Birkby}, {Brandl},
  {Brogi}, {Keller}, {Kenworthy}, {Schwarz}, \& {Stuik}}]{Snellen2015}
{Snellen}, I., {de Kok}, R., {Birkby}, J.~L., {et~al.} 2015, \aap, 576, A59

\bibitem[{{Stalport} {et~al.}(2023){Stalport}, {Cretignier}, {Udry}, {John},
  {Wilson}, {Delisle}, {Bonomo}, {Buchhave}, {Charbonneau}, {Dalal}, {Damasso},
  {Di Fabrizio}, {Dumusque}, {Fiorenzano}, {Harutyunyan}, {Haywood}, {Latham},
  {L{\'o}pez-Morales}, {Lorenzi}, {Lovis}, {Malavolta}, {Molinari}, {Mortier},
  {Pedani}, {Pepe}, {Pinamonti}, {Poretti}, {Rice}, \&
  {Sozzetti}}]{Stalport2023}
{Stalport}, M., {Cretignier}, M., {Udry}, S., {et~al.} 2023, \aap, 678, A90

\bibitem[{{Stock} {et~al.}(2020){Stock}, {Nagel}, {Kemmer}, {Passegger},
  {Reffert}, {Quirrenbach}, {Caballero}, {Czesla}, {B{\'e}jar}, {Cardona},
  {D{\'\i}ez-Alonso}, {Herrero}, {Lalitha}, {Schlecker}, {Tal-Or},
  {Rodr{\'\i}guez}, {Rodr{\'\i}guez-L{\'o}pez}, {Ribas}, {Reiners}, {Amado},
  {Bauer}, {Bluhm}, {Cort{\'e}s-Contreras}, {Gonz{\'a}lez-Cuesta}, {Dreizler},
  {Hatzes}, {Henning}, {Jeffers}, {Kaminski}, {K{\"u}rster}, {Lafarga},
  {L{\'o}pez-Gonz{\'a}lez}, {Montes}, {Morales}, {Pedraz}, {Sch{\"o}fer},
  {Schweitzer}, {Trifonov}, {Zapatero Osorio}, \& {Zechmeister}}]{Stock20}
{Stock}, S., {Nagel}, E., {Kemmer}, J., {et~al.} 2020, \aap, 643, A112

\bibitem[{{Su{\'a}rez Mascare{\~n}o} {et~al.}(2016){Su{\'a}rez Mascare{\~n}o},
  {Rebolo}, \& {Gonz{\'a}lez Hern{\'a}ndez}}]{SuarezMascareno2016}
{Su{\'a}rez Mascare{\~n}o}, A., {Rebolo}, R., \& {Gonz{\'a}lez Hern{\'a}ndez},
  J.~I. 2016, \aap, 595, A12

\bibitem[{{Su{\'a}rez Mascare{\~n}o} {et~al.}(2018){Su{\'a}rez Mascare{\~n}o},
  {Rebolo}, {Gonz{\'a}lez Hern{\'a}ndez}, {Toledo-Padr{\'o}n}, {Perger},
  {Ribas}, {Affer}, {Micela}, {Damasso}, {Maldonado}, {Gonz{\'a}lez-Alvarez},
  {Leto}, {Pagano}, {Scandariato}, {Sozzetti}, {Lanza}, {Malavolta}, {Claudi},
  {Cosentino}, {Desidera}, {Giacobbe}, {Maggio}, {Rainer}, {Esposito},
  {Benatti}, {Pedani}, {Morales}, {Herrero}, {Lafarga}, {Rosich}, \&
  {Pinamonti}}]{SuarezMascareno2018}
{Su{\'a}rez Mascare{\~n}o}, A., {Rebolo}, R., {Gonz{\'a}lez Hern{\'a}ndez},
  J.~I., {et~al.} 2018, \aap, 612, A89

\bibitem[{{Turbet} {et~al.}(2023){Turbet}, {Fauchez}, {Leconte}, {Bolmont},
  {Chaverot}, {Forget}, {Millour}, {Selsis}, {Charnay}, {Ducrot}, {Gillon},
  {Maurel}, \& {Villanueva}}]{Turbet2023}
{Turbet}, M., {Fauchez}, T.~J., {Leconte}, J., {et~al.} 2023, \aap, 679, A126

\bibitem[{{Wilson}(1968)}]{Wilson1968}
{Wilson}, O.~C. 1968, \apj, 153, 221

\bibitem[{{Zechmeister} \& {K{\"u}rster}(2009)}]{Zechmeister09}
{Zechmeister}, M. \& {K{\"u}rster}, M. 2009, \aap, 496, 577

\end{thebibliography}

\end{document}